\documentclass[12pt]{article}
\usepackage{a4wide,amssymb,cite}
\pdfoutput=1

\usepackage{a4wide,amssymb,graphicx}
\usepackage{epsfig}
\usepackage[usenames,dvipsnames]{color}
\usepackage{slashed}
\usepackage{amssymb,cite,graphicx}
\usepackage{slashed}
\usepackage{amsmath,bm,bbm}
\usepackage{amsfonts}
\usepackage[titletoc,title]{appendix}
\usepackage[small]{caption}
\usepackage[margin=1in]{geometry}
\usepackage[multiple]{footmisc}
\usepackage{mathtools}
\usepackage{slashed}
\usepackage[nottoc]{tocbibind}
\usepackage{xcolor}

\newcommand{\be}{\begin{equation}}
\newcommand{\ee}{\end{equation}}
\newcommand{\bea}{\begin{eqnarray}}
\newcommand{\eea}{\end{eqnarray}}

\def\circa#1{\,\raise.3ex\hbox{$#1$\kern-.75em\lower1ex\hbox{$\sim$}}\,}

\begin{document}

\begin{titlepage}

\rightline{CERN-TH-2020-087}

\begin{centering}
\vspace{1cm}
{\Large {\bf The Clockwork Standard Model}} \\

\vspace{1.5cm}

{\bf Yoo-Jin Kang$^1$, Soonbin Kim$^1$ and Hyun Min Lee$^{\dagger,1,2}$ }
\vspace{.5cm}

{\it $^1$Department of Physics, Chung-Ang University, Seoul 06974, Korea.} 
\\ \vspace{0.2cm}
{\it $^2$CERN, Theory department, 1211 Geneva 23, Switzerland. }

\end{centering}
\vspace{2cm}

\begin{abstract}
\noindent
 We consider various bulk fields with general dilaton couplings in the linear dilaton background in five dimensions as the continuum limit of clockwork models.  We show that the localization of the zero modes of bulk fields and the mass gap in the KK spectrum depend not only on the bulk dilaton coupling, but also on the bulk mass parameter in the case of a bulk fermion.
The consistency from universality and perturbativity of gauge couplings constrain the dilaton couplings to the brane-localized matter fields as well as the bulk gauge bosons. Constructing the Clockwork Standard Model (SM) in the linear dilaton background, we provide the necessary conditions for the bulk mass parameters for explaining the mass hierarchy and mixing for the SM fermions. We can introduce a sizable expansion parameter $\varepsilon=e^{-\frac{2}{3} kz_c}$ for the realistic flavor structure in the quark sector without a fine-tuning in the bulk mass parameters, but at the expense of a large 5D Planck scale. On the other hand, we can use a smaller expansion parameter for lepton masses, in favor of the solution to the hierarchy problem of the Higgs mass parameter. 
We found that massive Kaluza-Klein (KK) gauge bosons and massive KK gravitons couple more strongly to light and heavy fermions, respectively, so there is a complementarity in the resonance researches for those KK modes at the LHC.

\end{abstract}

\vspace{3cm}

 \begin{flushleft} 
$^\dagger$Email: hminlee@cau.ac.kr 
 \end{flushleft}

\end{titlepage}

\section{Introduction}

The Standard Model (SM) of particles physics has been tested well with precision experiments and the consistency of its inner structure has been confirmed with the discovery of the Higgs boson and new data from the Large Hadron Collider (LHC).
Gauge symmetry principle and quantum field theory, which are the core concepts for the SM, have provided the important guideline for extending the structure of the SM beyond the scales of proximity to accessible energies in the current experiments.
On the other hand, General Relativity (GR) has provided the crucial tools for explaining the cosmological history in the early Universe and all the way to the terrestrial phenomena of gravitation, becoming a new arena for testing physics beyond the SM, after the discovery of gravitational waves from the mergers of binary black holes at LIGO.  

The hierarchy problem and the flavor problem in the SM call for new physics beyond the SM, so new solutions to those problems have been main drivers for motivating direct and indirect experimental programs in the last decades. Solutions to the hierarchy problem require new particles and new symmetries close to the weak scale such as weak-scale supersymmetry, composite Higgs models, large or warped extra dimensions, etc,  but there have been no convincing hints for them even after ten years of running at the LHC.
Therefore, new ideas for solving the hierarchy problem \cite{hmlee} without new light colored partners of the SM at the weak scale have been suggested, such as neutral naturalness \cite{neutral}, relaxation mechanism \cite{crelax,relaxation}, clockwork models \cite{4DCW}, four-form flux models \cite{crelax,fourform}, etc.

The clockwork models have drawn new attention as the solution to the hierarchy problems in particles physics, not necessarily related to the hierarchy problem of the Higgs mass parameter. 
The idea is based on the multiple copies of particles or symmetries with nearest neighbor interactions in four dimensions, explaining the small couplings from the localization of the lightest mode in the theory space even for order one coupling of each copy.
There is a counterpart of the continuum limit of clockwork models in the linear dilaton background in five dimensions, where the zero mode of a bulk field has position-dependent couplings in the extra dimension due to the warped factor.
There is a mass gap between the zero mode and a stack of massive Kaluza-Klein (KK) modes in clockwork models and the continuum counterpart, becoming a smoking-gun signal at the LHC.

In this article, we consider various bulk fields with general dilaton couplings in the linear dilaton background in five dimensions, so called the clockwork gravity, as an extension of the GR, and discuss the particle spectrum and the effective couplings in each case in the 4D effective theory as the continuum limit of clockwork models \cite{4DCW,CWphoton,CW5d,CWgraviton,CWsugra,CWp2}. 
The hierarchy problem in the clockwork gravity can be solved due to the delocalization of the massless graviton away from the brane where the Higgs field is localized.  
From the general discussion on bulk fields including scalars, fermions, gauge bosons as well as graviton, we identify the effective couplings between the zero mode of matter fields and the massive KK modes of gauge bosons and graviton.

Based on the results of our general discussion, we construct the ``Clockwork Standard Model'' where all the SM particles except the Higgs field propagate into the bulk.
Investigating the flavor structure of quarks and leptons from the localization of chiral zero modes in this construction, we study the implications of the Clockwork SM as the simultaneous solutions to the hierarchy problem and the flavor problem, and discuss the possibility for complementary searches of massive KK modes at the LHC. 

For the Randall-Sundrum (RS) warped background, a similar construction of the bulk SM has been discussed in the literature \cite{bulkRS} and the flavor issues associated with the bulk SM were thoroughly investigated \cite{rsnu,rsflavor,rsflavor2}.  
In the case of the linear dilaton background,  flavor or dark matter puzzles were also studied mainly in the context of four-dimensional clockwork models \cite{cwflavor,cwnu,cwwimp,general,CWphoton,CWg2}.

The paper is organized as follows.
We begin with a brief review on the gravitational action with a dilaton in five dimensions and discuss  the linear dilaton background in connection to the solution to the hierarchy problem of the Higgs mass parameter.
Then, we introduce bulk scalars with bulk mass only in Jordan frame and discuss the profiles of the zero mode as well as the massive modes. We continue to extend our analysis to bulk fermions with bulk mass parameters of kink type and show the nontrivial profiles of localization of the chiral zero mode, depending on the dilaton coupling and the bulk mass parameters.
When bulk matter fields carry gauge charges, it is necessary to let the corresponding gauge bosons propagate into the bulk. Thus, we also introduce bulk gauge bosons and identify the consistent dilaton couplings required for universality and perturbavitivity as well as compute the couplings between the zero modes of matter fields and the massive KK modes of gauge bosons. Next, we show the general couplings of massive KK gravitons to the brane fields as well as the zero modes of bulk fields.

Putting the pieces of the obtained results together, we next construct the Clockwork Standard Model and discuss the mass hierarchy and mixing for quarks and leptons from the overlaps between the zero modes of matter fields and show the implications for the effective couplings of the SM particles to the massive KK gauge bosons and gravitons.
Finally, conclusions are drawn. There is one appendix dealing with the transformations of the Lagrangians for bulk and brane matter fields from Jordan to Einstein frames.

\section{The Clockwork Gravity}

We first review the warped geometry with the linear dilaton background in five dimensions.
The model is the counterpart of the continuum Clockwork(CW) models in four dimensions where there are multiple copies of identical particles or symmetries with nearest neighbor interactions \cite{4DCW,CWphoton,CWgraviton}.

The five-dimensional gravity action with a dilaton $S$ in Jordan (or string) frame \cite{CW,CW2,CWradion,CWgraviton} is given by
\bea
S&=&\int d^5x \sqrt{G}\, \frac{M^3_5}{2} \, e^S \Big( -R(G)-(\partial_M S)^2 + 4k^2 \Big) \nonumber \\
&&- \int d^5x \sqrt{G} \,e^S \Big(\frac{\delta(z)}{\sqrt{-G_{55}}}\,\Lambda_1+\frac{\delta(z-z_c)}{\sqrt{-G_{55}}}\,\Lambda_2 \Big) \label{jordan}
\eea
where $M_5$ is the 5D Planck mass, $k$ is the 5D curvature and $\Lambda_{1,2}$ are the brane tensions.  Then, the scale symmetry with $S\rightarrow S+\delta$ and $G_{MN}\rightarrow e^{-2\delta/3}G_{MN} $ is broken explicitly by a nonzero $k^2$ as well as brane tensions. 
With a Weyl rescaling of the metric, $G_{MN}=e^{-2S/3} G^E_{MN}$, the above action becomes in Einstein frame
\bea
S&=& \int d^5x \sqrt{G_E}\, \frac{M^3_5}{2}\, \Big(-R(G_E)+\frac{1}{3}(\partial_M S)^2+ 4k^2\, e^{-\frac{2}{3}S}\Big) \nonumber \\
&&- \int d^5x \sqrt{G_E} \,e^{-S/3} \Big(\frac{\delta(z)}{\sqrt{-G^E_{55}}}\,\Lambda_1+\frac{\delta(z-z_c)}{\sqrt{-G^E_{55}}}\,\Lambda_2 \Big). 
\eea
Then, in Einstein frame, there appear a dilaton potential in the bulk and dilaton-dependent couplings to the branes.

Then, there exists a warped solution to the bulk Einstein equation, whose metric satisfying the $Z_2$ symmetry, $z\rightarrow -z$, is given by
\bea
ds^2=(w(z))^2 (\eta_{\mu\nu} dx^\mu dx^\nu - dz^2) \label{CWmetric}
\eea
with
$w(z)=e^{\frac{2}{3}k|z|}$, together with  linear dilaton background, $S(z)=2k|z|$. The metric in eq.~(\ref{CWmetric}) can be rewritten in another coordinate $y=\int \omega(z) dz$ as
\bea
ds^2 = e^{-2\sigma(y)}\eta_{\mu\nu} dx^\mu dx^\nu -dy^2, 
\eea
with  $\sigma=-\ln \Big(\frac{2}{3} k|y|+1\Big)$ and $S(y)=3\ln \Big(\frac{2}{3} k|y|+1\Big)$  where we chose the warp factor to unity at $y=0$ without a loss of generality. Thus, we denote $\omega^2=e^{-2\sigma}=e^{\frac{2}{3}S}$. 
Then, the extra dimension is bounded to $z\in(-z_c,z_c]$ or $y\in(-y_c,y_c]$, and the  tuning relations between brane tensions in the linear dilaton background are required for the consistency of the warped metric solution at the branes, as follows,
\be
\Lambda_2=-\Lambda_1=4k M^3_5. 
\ee
Fom the relation between Jordan and Einstein frame metrics,  $G_{MN}=e^{-2S/3} G^E_{MN}=\omega^{-2} G^E_{MN}$, we note that the Jordan frame metric becomes nothing but the 5D Minkowski spacetime, namely, $G_{MN}=\eta_{MN}$.

The effective 4D Planck mass for the  dilaton background is also given by
\bea
M^2_P=M^3_5 \int^{z_c}_{-z_c} dz\,  w^3 =M^3_5 \int^{z_c}_{-z_c} dz\,  e^{2 k|z|}= \frac{M^3_5}{k}\, \Big(e^{2kz_c} -1\Big). \label{Pmass-CW}
\eea
In terms of the proper length of the extra dimension, 
\bea
L_5=\int^{z_c}_{-z_c} dz\, w=\int^{z_c}_{-z_c} dz\, e^{\frac{2}{3}k|z|}= \frac{3}{k} \, \Big(e^{\frac{2}{3}kz_c} -1\Big)=2y_c. \label{length}
\eea
we can rewrite the 4D Planck mass (\ref{Pmass-CW}) as
\bea
M^2_P\approx \frac{1}{27} M^3_5 k^2 L^3_5. \label{approx}
\eea
Therefore, the proper length of the extra dimension $L_5$ can be much larger than the inverse of the 5D curvature scale $k$ due to the exponential factor, $e^{\frac{2}{3}kz_c}$, so the result can make one warped extra dimension and a small 5D Planck mass compatible with the phenomenological constraints, unlike the case with one flat extra dimension \cite{ADD}. From eq.~(\ref{approx}), the relation between the 4D Planck scale and the length of the extra dimension looks like the toroidal compactification of a 7D gravity.
As a result, the linear dilaton background differs from the warped extra dimension without a dilaton \cite{RS}, because the 5D Planck scale can be taken to a small value due to the exponentially large proper length of the extra dimension in the former case.
In the later discussion, we use the exponential warp factor in the conformal coordinate $z$ and the large proper length in the Gaussian normal coordinate $y$ interchangeably by keeping in mind the relation in eq.~(\ref{length}). 

In the linear dilaton background, it is remarkable that $M_5$ and $k<M_5$ can be much smaller than the Planck scale, thus addressing the hierarchy problem with the warped extra dimension and allowing for the KK masses of order $k$ as will be discussed in the later sections. 
From eq.~(\ref{Pmass-CW}),  we obtain the condition for the 5D curvature scale to satisfy
\bea
kz_c+\frac{1}{2}\ln\Big(1-e^{-2kz_c}\Big)=32+\frac{1}{2} \ln  \Big(\frac{k}{1\,{\rm TeV}} \Big) -\frac{3}{2} \ln  \Big(\frac{M_5}{10\,{\rm TeV}} \Big). \label{hierarchy}
\eea
Therefore, for the large 4D Planck mass or weak gravity, we only need a mild hierarchy between the 5D Planck mass and the electroweak scale in the model with one extra dimension, thanks to the warp factor with $kz_c\sim 30$. In this case, a mild hierarchy between the 5D curvature scale $k$ and the radius of the extra dimension $z_c$ can be guaranteed by the stabilization mechanism for the dilaton field $S$ \cite{CWradion}.  For a small $M_5$, the expansion parameter becomes $\varepsilon=e^{-\frac{2}{3}k z_c}\simeq 5\times 10^{-10}$, which would be appropriate for explaining the smallness of neutrino masses, as will be discussed in the later section. 

Having in mind the application of the clockwork framework to the flavor problems in the SM,  we also rewrite eq.~(\ref{hierarchy}) for a large $M_5$, as follows,
\bea
kz_c+\frac{1}{2}\ln\Big(1-e^{-2kz_c}\Big)=3.2+\frac{1}{2} \ln  \Big(\frac{k}{10^5\,{\rm TeV}} \Big) -\frac{3}{2} \ln  \Big(\frac{M_5}{10^{11}\,{\rm TeV}} \Big).
\eea
Then, in the case with a large $M_5$, we can take a smaller value of $kz_c$ in the warp factor such that the expansion parameter becomes $\varepsilon=e^{-\frac{2}{3}k z_c}\simeq 0.12$, which is appropriately sizable for obtaining the realistic quark Yukawa couplings from the bulk fermions in the later discussion.

\section{Bulk Scalars}

The Lagrangian for a bulk real scalar field $\chi$ with the general dilaton coupling in Jordan frame \cite{4DCW}  is given by
\bea
{\cal L}_S= \sqrt{G}\, e^{c S}\, \Big(\frac{1}{2}  G^{MN}\partial_M\chi \partial_N\chi-\frac{1}{2}m^2_\chi \chi^2 \Big)  \label{scalar}
\eea 
where $c$ is a constant parameter for the dilaton coupling.
In the case where a bulk complex scalar field is charged under a gauge symmetry with bulk gauge boson $A_M(x,z)$, we can promote the derivative in the above Lagrangian to a covariant derivative as $D_M=\partial_M-i q_\chi g_{5D}  A_M$ with $g_{5D}$ being the 5D gauge coupling and $q_\chi$ being the charge of the complex scalar field $\chi$. We can also add a bulk potential as well as a brane potential, being consistent with the symmetry.

 Then, the Euler equation for the Lagrangian (\ref{scalar}) is
 \bea
 \frac{1}{\sqrt{G}} \, \partial_M \Big(\sqrt{G} \, e^{c S}\,  G^{MN} \partial_N \chi \Big)+  e^{c S} m^2_\chi \chi=0. \label{scalar-jordan}
 \eea

 On the other hand, in Einstein frame with $G_{MN}=e^{-2S/3} G^E_{MN}$, we have the original Lagrangian (\ref{scalar}) in the following form,
 \bea
{\cal L}_S= \sqrt{G_E}\, \Big(\frac{1}{2}\, e^{(c-1)S}\, G^{MN}_E\partial_M\chi \partial_N\chi -\frac{1}{2}\, e^{(c-\frac{5}{3})S}\, m^2_\chi \chi^2 \Big). 
\eea 
Then, the corresponding Euler equation is
\bea
 \frac{1}{\sqrt{G_E}} \, \partial_M \Big(\sqrt{-G_E} \,  e^{(c-1)S}\, G^{MN}_E \partial_N \chi \Big)+  e^{(c-\frac{5}{3})S} m^2_\chi \chi=0.
 \eea

\subsection{Scalar clockwork modes} 
 
For the flat metric in Jordan frame, $G_{MN}=\eta_{MN}$, the equation of motion for the bulk scalar in eq.~(\ref{scalar-jordan}) becomes
\bea
\Box \chi -c S' \partial_z\chi - \partial^2_z\chi + m^2_\chi \chi=0.  \label{CWscalar}
\eea
Then, taking $\chi=e^{- kc|z|}\, \sum_n \chi^{(n)}(x) f^\chi_n(z)$ with $(\Box+m^2_n) \chi^{(n)}(x)=0$, we can cast the above equation into the equation for the mode function $f^\chi_n(z)$,
\bea
\partial^2_z f^\chi_n+ (m^2_n-k^2c^2-m^2_\chi) f^\chi_n-2kc\big( \delta(z)-\delta(z-z_c) \big) f^\chi_n =0.
\eea 
Here, there appear Dirac delta terms due to the second derivatives of  the $Z_2$ symmetric factor in the redefined field.

First, for the zero mode solution, we have $f^\chi_0\propto e^{\pm \sqrt{k^2c^2+m^2_\chi} z}$, but there is no zero mode solution satisfying the Neumann boundary conditions, $\partial_z \chi=0$ at $z=0$ and $z_c$, unless $m_\chi=0$, as expected for a massive bulk scalar. 
Taking $m_\chi=0$,  a constant zero mode solution for $\chi$ exists,
\bea
f^\chi_0=N_{\chi_0}\,  e^{ kc  |z|} \label{szero}
\eea
with
\bea
 N_{\chi_0}= \sqrt{\frac{kc }{e^{2kc z_c}-1}} \label{bnorm}
 \eea
Here, the normalization factor $N_{\chi_0}$ is determined by $2\int^{z_c}_0 dz\, (f^\chi_0)^2=1$. In this case, the wave functions for massive modes are given by
\bea
f^\chi_n = N_{\chi_n} \Big(\cos\frac{\pi n z}{z_c} + \frac{kc z_c}{\pi n}\sin\frac{\pi n |z|}{z_c} \Big), \quad n\in Z,
\eea
with the eigenvalues and normalization factors being
  \bea
 m^2_{\chi_n}&=& c^2 k^2 + \frac{\pi^2 n^2}{z^2_c}, \\
 N_{\chi_n} &=&\frac{1}{\sqrt{z_c}}\, \Big(\frac{\pi n}{z_c m_{\chi_n}}\Big).  \label{bnormkk}
 \eea
 Therefore, we find that the KK masses depend on the dilaton coupling $c$  and the 5D  curvature scale $k$ as well as the radius of the extra dimension $z_c$. Taking $c=1$, we can recover the results as a continuum limit of the 4D clockwork scalars. 
However, for $c\neq 1$, the mass gap between the zero mode and the first KK mode is given by $c\,k$. 

We note that in the Gaussian normal coordinate $y$, related to the conformal coordinate $z$ by $dy=e^{\frac{2}{3}kz}\, dz$, the zero mode solution for the bulk scalar field in eq.~(\ref{szero}) becomes
$f^\chi_0=N_{\chi_0}\, (\frac{2}{3}k|y|+1)^{\frac{3}{2}c}$ with $ N_{\chi_0}=\sqrt{kc/[(\frac{1}{3} kL_5+1)^{3c}-1]}$. 
Then, in the $y$ coordinate, the exponential warp factor is replaced by the power-law factor, but instead with a large proper length, $L_5$, in the normalization factor.
But, henceforth, we keep on working in the conformal coordinate $z$ for convenience.

On the other hand, for $m_\chi\neq 0$, the eigenvalues and normalizations for massive modes become
 \bea
 m^2_{\chi_n}&=& m^2_\chi + c^2 k^2  + \frac{\pi^2 n^2}{z^2_c}, \\
 N_{\chi_n} &=&\frac{1}{\sqrt{z_c}}\,\frac{\pi n}{z_c \sqrt{m^2_{\chi_n}-m^2_\chi}}. 
 \eea
 Therefore, the squared masses for the KK modes of the massive bulk scalar field are shifted by the bulk mass.

\subsection{Localized couplings of scalar clockwork}

Suppose that a massless bulk scalar couples to the external operators localized on the branes.
Then, the effective couplings in four dimensions depend on the mode function of  the scalar as well as the dilaton coupling. For instance, we can introduce the scalar coupling to the external operators ${\cal O}_{1, {\rm ext}},{\cal O}_{2, {\rm ext}}$ localized at $z=0, z_c$ in Jordan frame,
\bea
{\cal L}_{S,{\rm int}} = \frac{\sqrt{G}}{\sqrt{-G_{55}}}\, e^{\frac{1}{2}S}\,\chi\Big(\delta(z)\,{\cal O}_{1,{\rm ext}}+ \delta(z-z_c)\,{\cal O}_{2,{\rm ext}} \Big). \label{scalarcoupling}
\eea
Then, for $c>0$ and $e^{k c z_c}\gg 1$,  the normalization factor becomes $N_{\chi_0}\simeq \sqrt{kc}\, e^{-k c z_c}$, so the effective coupling to the zero mode $\chi^{(0)}$ is exponentially suppressed at $z=0$, but not at  $z=z_c$.  We note that the couplings to the KK scalars  $\chi^{(n)}$ at either branes are of similar order at both branes. For instance, an axion-like scalar field can be introduced  in the bulk with the brane-localized coupling to the SM gluons by ${\cal O}_{1,{\rm ext}}=\frac{\alpha}{8\pi f_a}\, G_{\mu\nu}{\tilde G}^{\mu\nu}$ at $z=0$.  In this case, the massless mode of the bulk axion has a large effective axion decay constant \cite{4DCW}, $f_{a,{\rm eff}}= e^{k c z_c}\, f_a \gg f_a$, below the KK mass scale.

We also note that even for $c<0$ and $e^{k |c| z_c}\gg 1$, the relative exponential suppression of the scalar coupling at $z=0$ as compared to $z=z_c$ is maintained, even though the normalization factor becomes $N_{\chi_0}\simeq \sqrt{k|c|}$. But, in this case, we need the overall suppression of the coefficients of the external operators for perturbativity at $z=z_c$.

\section{Bulk Fermions}

We consider a bulk fermion with the dilaton factor in Jordan frame, similarly to the case with a bulk scalar in eq.~(\ref{scalar}), as follows,
\bea
{\cal L}_F = \sqrt{G} \, e^{c S}\bigg[ i{\bar\psi} \Gamma^M\Big( \partial_M +\frac{1}{8} \omega_M\,^{\underline{A}\underline{B}}[\Gamma_{\underline{A}},\Gamma_{\underline{B}}]  \Big)\psi-e^{\frac{1}{3}S}\,m(y){\bar\psi}\psi  \bigg].
\eea
Here, the dilaton factor $e^{cS}$ can be in principle different from the one for the bulk scalar discussed in the previous section.
But, in the later discussion on the clockwork SM, we will assume that the bulk matter fields take the same dilaton couplings.

In the case where the bulk fermion is charged under a gauge symmetry with bulk gauge boson $A_M(x,z)$, we can promote the derivative to a covariant derivative as $D_M=\partial_M-i q_\psi g_{5D}  A_M$ with $g_{5D}$ being the 5D gauge coupling and $q_\chi$ being the charge of the field $\psi$.
Here, the bulk fermion mass is given by $m(y)=\nu \sigma'$ where $\nu$ the bulk mass parameter. We have $\sigma'=-\frac{2}{3} k\, e^\sigma\,{\rm sgn}(y)=-\frac{2}{3} k\, e^{-\frac{1}{3}S}\,{\rm sgn}(y)$, so $e^{\frac{1}{3}S} m(y)=-\frac{2}{3} k\,{\rm sgn}(y)$, resulting the constant  bulk mass term in Jordan frame, except at the branes.

Now going to the Einstein frame with $G_{MN}=e^{-2S/3} G^E_{MN}$, we can rewrite the above Lagrangian for the bulk fermion as 
\bea
{\cal L}_F = \sqrt{G_E} \, e^{cS} \bigg[i{\bar\psi}' \Gamma^M\Big( \partial_M +\frac{1}{8} \omega_M\,^{\underline{A}\underline{B}}[\Gamma_{\underline{A}},\Gamma_{\underline{B}}]  \Big)\psi'-m(y){\bar\psi}'\psi'  \bigg] \label{bulkf2}
\eea
where we have rescaled the bulk fermion by $\psi'=e^{-2S/3}\psi$ in order to make the covariant derivative invariant.
This is similar to the bulk fermion Lagrangian considered for the bulk RS model, except the dilaton factor \cite{bulkRS}.
Here, $\Gamma^M=\Gamma^{\underline{A}}\, e_{\underline{A}}\,^M$, $\Gamma^{\underline{A}}=(\gamma_\mu,i\gamma_5)$,  $\Gamma_{\underline{A}}=(\gamma_\mu,-i\gamma_5)$, and $\{\Gamma^{\underline{A}},\Gamma^{\underline{B}}\}=2\eta^{\underline{A}\,\underline{B}}=2\,{\rm diag}(+,-,-,-,-)$. The vielbein $e_{\underline{A}}$ relates the curved metric to the flat metric by $e_{\underline{A}}\,^M e_{\underline{B}}\,^N G_{E,MN} =\eta_{\underline{A} \underline{B}}$. 
The spin connection in 5D is defined as
\bea
\omega_{M\underline{A}\,\underline{B}} = \frac{1}{2}\Big(e_{\underline{A}}\,^P \Omega_{MP\underline{B}}-e_{\underline{B}}\,^P \Omega_{MP\underline{A}}-e_{\underline{A}}\,^P e_{\underline{B}}\,^Q  e^{\underline{C}}\,_M \Omega_{PQ\underline{C}} \Big) 
\eea
with $\Omega_{MN\underline{A}}=\partial_M e_{N\underline{A}}-\partial_N e_{M\underline{A}}$.
For the warped metric in Einstein frame, $G_{E,MN}={\rm diag}(e^{-2\sigma},-e^{-2\sigma},-e^{-2\sigma},-e^{-2\sigma},-1)$, given in eq.~(\ref{CWmetric}), the nonzero components of the vielbein  are $e_{\underline{\alpha}}\,^\mu=e^\sigma\, \delta^\mu_{\underline{\alpha}}$ and $e_{\underline{5}}\,^5=1$, so the only nonzero components of the spin connection are
\bea
\omega_\mu\,^{\underline{\alpha}\,\underline{5}}=\sigma'\, e^{-\sigma} \delta_\mu^{\underline{\alpha}}.
\eea 
Then, with $e^{cS}=e^{-3c\sigma}$, the bulk fermion Lagrangian (\ref{bulkf2}) is further simplified to
\bea
{\cal L}_F &=& e^{-3(c+1)\sigma} {\bar\psi}' \bigg[i\gamma^\mu\partial_\mu -\gamma_5 \, e^{-\sigma}\,\partial_5 + e^{-\sigma}\sigma' \Big(2\gamma_5-\nu\Big) \bigg]\psi' \nonumber \\
&=&\overline{{\widetilde\psi}} \bigg[i\gamma^\mu\partial_\mu -\gamma_5 \, e^{-\sigma}\,\partial_5 + e^{-\sigma}\sigma' \Big(\frac{1}{2}(1-3c)\gamma_5-\nu \bigg) \Big]{\widetilde\psi} \label{fermion}
\eea
with
\bea
{\widetilde\psi} \equiv e^{-\frac{3}{2}(c+1)\sigma} \psi'.
\eea
As a result, the Euler equation for the redefined fermion is
\bea
i\gamma^\mu\partial_\mu{\widetilde\psi} -\gamma_5 \, e^{-\sigma}\,\partial_5{\widetilde\psi} + e^{-\sigma}\sigma' \Big(\frac{1}{2}(1-3c)\gamma_5-\nu \bigg){\widetilde\psi}=0.
\eea

\subsection{Fermion clockwork modes}

We impose the boundary condition on the bulk fermion by 
\bea
\gamma_5 \psi(x,-y)= -\psi(x,y).
\eea
Then, only the left-handed fermion has a massless mode by
\bea
{\widetilde\psi}(x,y)= N_{L}\, \psi^{(0)}_L (x) f_{L}(y), \qquad f_{L}(y)= e^{\big(\frac{1}{2}(1-3c)+\nu\big)\sigma}, \label{fermion1}
\eea
where $i\gamma^\mu\partial_\mu \psi^{(0)}_L (x)=0$ with $\gamma_5  \psi^{(0)}_L (x)=- \psi^{(0)}_L (x)$, 
and $N_{L}$ is the normalization factor, determined by $2N^2_{L}\int^L_0 dy (f_{L}(y))^2=1$. 
If we impose an alternative boundary condition on the bulk fermion by
\bea
\gamma_5 \psi(x,-y)= \psi(x,y),
\eea
only the right-handed fermion has a massless mode by
\bea
{\widetilde\psi}(x,y)= N_{R}\, \psi^{(0)}_R (x) f_{R}(y), \qquad f_{R}(y)= e^{\big(\frac{1}{2}(1-3c)-\nu\big)\sigma}, \label{fermion2}
\eea
where $i\gamma^\mu\partial_\mu \psi^{(0)}_R (x)=0$ with $\gamma_5  \psi^{(0)}_R (x)= \psi^{(0)}_R (x)$, 
and $N_{R}$ is also determined by 
$2N^2_{R}\int^L_0 dy (f_{R}(y))^2=1$.
With the normalizations in the $z$ coordinate, $2N^2_{L,R}\int^{z_c}_0 dz \,( e^{-\frac{1}{2}\sigma}\, f_{L,R}(z))^2=1$,  the zero mode wave functions as the probability densities are given by
\bea
e^{-\frac{1}{2}\sigma}\, f_{L,R}(z)=e^{\frac{1}{3}(3c\mp 2\nu)k|z|}, \label{fzero}
\eea
and the normalization factors are
\bea
N_{L,R} =\sqrt{\frac{\frac{1}{3}(-3c\pm2\nu) k}{ 1- e^{-\frac{2}{3}(-3c\pm2 \nu) k z_c}}}. \label{fnorm}
\eea
Then, we find the localization behavior of the fermion zero modes with respect to the fixed point $z=0$: for $\nu>\frac{3}{2}c\, (\nu<\frac{3}{2}c)$, the left-handed zero mode is localized at $z=0 (z=z_c)$; for $\nu>-\frac{3}{2}c\, (\nu<-\frac{3}{2}c)$, the right-handed zero mode is localized at $z=z_c (z=0)$.
Therefore, if $|\nu|<\frac{3}{2} c$ for $c>0$ or $|\nu|>\frac{3}{2}|c|$ for $c<0$, either of chiral zero modes are localized towards $z=z_c$.
If $\nu=\pm \frac{3}{2}c$ for the zero mode of a left-handed or right-handed fermion,  we obtain the normalization factors as $N_{L,R}=\frac{1}{\sqrt{2z_c}}$, which does not depend on the warp factor. 
 
As in the bulk scalar case in the previous section, in the Gaussian normal coordinate $y$, the zero mode solutions for the bulk fermion in eqs.~(\ref{fermion1}) and (\ref{fermion2}) become $f_{L,R}=(\frac{2}{3}k|y|+1)^{-(\frac{1}{2} (1-3c)\pm \nu)}$ with $N_{L,R}=\sqrt{\frac{1}{3}(-3c\pm2\nu) k/\Big(1-(\frac{1}{3}kL_5+1)^{(3c\mp 2\nu)}\Big)}$. Thus, in the $y$ coordinate, the exponential factor is replaced by the power-law factor, but with a large proper length of the extra dimension in the normalization factors.
As will be discussed in the later sections, the large hierarchy of the effective Yukawa couplings in 4D can be attributed to the localization in the extra dimension with a large proper length. 

Taking the KK decomposition of the bulk fermion as 
\bea
\psi'(x,y) = \sum_n \Big[\psi^{(n)}_L(x)\xi_n(y)+ \psi^{(n)}_R(x)  \eta_n(y)\Big],
\eea
the fermion Lagrangian (\ref{fermion}) becomes
\bea
{\cal L}_F= \sum_n \Big[{\bar\psi}^{(n)}_L i\gamma^\mu \partial_\mu \psi^{(n)}_L+{\bar\psi}^{(n)}_R i\gamma^\mu \partial_\mu \psi^{(n)}_R -\Big(m_{\psi_n}{\bar\psi}^{(n)}_L \psi^{(n)}_R +{\rm h.c.}  \Big) \Big].
\eea
Here, we used the normalizations,
\bea
\int dy\, e^{-3(c+1)\sigma} \xi_n(y) \xi_m(y) = \int dy\, e^{-3(c+1)\sigma} \eta_n(y) \eta_m(y) =\delta_{mn},
\eea
and the wave functions, ${\widetilde\xi}_n =e^{-\frac{3}{2}(c+1)\sigma }\xi_n$ and ${\widetilde\eta}_n =e^{-\frac{3}{2}(c+1)\sigma }\eta_n$,  satisfy
\bea
 -e^{-\sigma}\Big[ \partial_5 - \Big(\frac{1}{2}(1-3c)+\nu \Big)\sigma'\Big] {\widetilde\xi}_n &=&  m_{\psi_n} {\widetilde\eta}_n,   \label{meq1}\\
  e^{-\sigma}\Big[ \partial_5 -\Big(\frac{1}{2}(1-3c)-\nu \Big)\sigma'\Big] {\widetilde\eta}_n &=& m_{\psi_n} {\widetilde\xi}_n.  \label{meq2}
\eea
Then, redefining with
\bea
{\widetilde\xi}_n = e^{\big(\frac{1}{2}(1-3c)+\nu\big)\sigma} \,\alpha_n, \quad {\widetilde\eta}_n = e^{\big(\frac{1}{2}(1-3c)-\nu\big)\sigma} \,\beta_n,
\eea
eqs.~(\ref{meq1}) and (\ref{meq2}) become
\bea
\partial_5 \alpha_n &=& - e^{(1-2\nu)\sigma} m_{\psi_n} \beta_n,  \label{mode1} \\
\partial_5 \beta_n &=& e^{(1+2\nu)\sigma} m_{\psi_n} \alpha_n. \label{mode2}
\eea
Consequently, we find that $\alpha_n$ satisfies the following second-order differential equation,
\bea
\Big[\partial^2_5 -(1-2\nu)\sigma' \partial_5 + m^2_{\psi_n}\, e^{2\sigma} \Big]\alpha_n=0.  \label{meq3}
\eea
Then, the other mode function, $\beta_n$, can be obtained from eq.~(\ref{mode2}), so it is sufficient to find the solutions for $\alpha_n$ from the above differential equation. But, for completeness, we also present the differential equation for $\beta_n$ as
\bea
\Big[\partial^2_5 -(1+2\nu)\sigma' \partial_5 + m^2_{\psi_n}\, e^{2\sigma} \Big]\beta_n=0.  \label{meq3b}
\eea

We note that for the boundary conditions, $\gamma_5 \psi(x,-y)=-\psi(x,y)$ and $\psi(x,y+2L)=\psi(x,y)$, and $\sigma(-y)=\sigma(y)$, we have ${\xi}_n(-y)=-{\xi}_n(y)$ or  the Neumann boundary conditions, $\partial_5 {\xi}_n(y=y_i)=0$  with $y_i=0,L$ on the orbifold. On the other hand, for $\gamma_5 \psi(x,-y)=\psi(x,y)$, we have ${\eta}_n(-y)={\eta}_n(y)$ or  the Dirichlet boundary conditions, ${\eta}_n(y=y_i)=0$ with $y_i=0,L$ on the orbifold. 

For $dy= e^{-\sigma} dz = e^{\frac{2}{3}kz} dz$, we can rewrite eq.~(\ref{meq3}) as
\bea
\Big( \frac{\partial^2}{\partial z^2}-\frac{4}{3} \nu k \frac{\partial}{\partial z}+m^2_{\psi_n}\Big) \alpha_n=0.
\eea
Then, for $a_n = e^{-\frac{2}{3}\nu k z}\, \alpha_n$, we have
\bea
\Big(  \frac{\partial^2}{\partial z^2}+ (m^2_{\psi_n}-\kappa^2) \Big) a_n=0, \quad \kappa= \frac{2}{3} \nu k,
\eea
that is, the general solution for $a_n$ is
\bea
a_n= A_n \cos\Big(\sqrt{m^2_{\psi_n}-\kappa^2}\, z\Big) +B_n \sin\Big(\sqrt{m^2_{\psi_n}-\kappa^2}\, z\Big).
\eea
Thus, for $\partial_5 {\xi}_n(y=y_i)=0$ with $y_i=0,L$, we have  the probability density functions of massive modes with $2\int_0^{z_c} dz (e^{-\frac{1}{2}\sigma}{\tilde\xi}_n)^2=1$, as follows, 
\bea
e^{-\frac{1}{2}\sigma}{\tilde\xi}_n&=&  e^{\big(-\frac{3}{2}c+\nu\big)\sigma} \, e^{\frac{2}{3}\nu k |z|} a_n
\nonumber \\
&=& \sqrt{\frac{1}{z_c}} \,e^{ck|z|} \, \cos\Big(\frac{n \pi z}{z_c}\Big),
 \label{fmassive}
\eea
and the mass eigenvalue is given by
\bea
m^2_{\psi_n}= \frac{n^2\pi^2}{z^2_c} + \frac{4}{9} \nu^2 k^2, \quad n\in Z. \label{kkfmass}
\eea
As a result, the mass gap between the zero mode and the first KK mode depends on the bulk mass parameter $\nu$ as well as the 5D curvature scale $k$, but not on the dilaton coupling $c$, unlike the case with a bulk scalar and a bulk gauge boson, as will be discussed in the next section.
In the limit of a strong localization of the zero mode with $|\nu|\gg 1$, we find that the KK fermions become decoupled, being consistent with the fact that the zero mode becomes a four-dimensional field.

Similarly, the other mode functions, $\eta_n$, have the same mass eigenvalues as in eq.~(\ref{kkfmass}), but the corresponding probability density functions, with $\int_0^{z_c} dz (e^{-\frac{1}{2}\sigma}{\tilde\eta}_n)^2=1$, satisfying the Dirichlet boundary conditions at $y_i=0, L$, are given by
\bea
e^{-\frac{1}{2}\sigma}{\tilde \eta}_n &=&  e^{\big(-\frac{3}{2}c-\nu\big)\sigma} \, e^{-\frac{2}{3}\nu k |z|} a_n \nonumber  \\
&=& \sqrt{\frac{1}{z_c}}\, e^{ck|z|} \, \sin\Big(\frac{n \pi |z|}{z_c}\Big).
\eea

\subsection{Localized couplings of fermion clockwork}

We consider a Yukawa coupling between the Higgs field localized at $z=0$ and two bulk fermions, $\psi$ and $\psi'$, with bulk mass parameters, $\nu_{\psi}$ and $\nu_{\psi'}$, respectively, 
\bea
{\cal L}_{\psi,{\rm int}}=- \delta(z)\, \frac{ \sqrt{G} }{\sqrt{-G_{55}}}\Big(y_\psi \,  {\bar \psi}_L \psi'_R H +{\rm h.c.} \Big).
\eea
Then, we require the left-handed or right-handed zero modes for $\psi$ and $\psi'$ to exist, that is, $\nu_\psi<\frac{3}{2}c$ and $\nu_{\psi'}>-\frac{3}{2}c$ for $c>0$. Thus, with $N_{\psi}\approx \sqrt{\frac{k}{3}(3c-2\nu_\psi)}\, \varepsilon^{\frac{3}{2}c- \nu_\psi}$ and $N_{\psi'}\approx \sqrt{\frac{k}{3}(3c+2\nu_{\psi'})}\, \varepsilon^{\frac{3}{2}c+\nu_{\psi'}}$ for $\varepsilon\equiv e^{-\frac{2}{3}kz_c}\ll 1$, we get the effective Yukawa coupling suppressed by the warp factor, as follows,
\bea
{\cal L}_{\psi,{\rm int}}&=&-y_\psi \, N_\psi N_{\psi'} {\bar \psi}_{L,0} \psi'_{R,0} H +{\rm h.c.}  \\
&=&-\lambda_\psi \,  {\bar \psi}_{L,0} \psi'_{R,0} H +{\rm h.c.}
\eea
with
\bea
\lambda_\psi \approx \frac{1}{3} k y_\psi\sqrt{(3c-2\nu_\psi)(3c+2\nu_{\psi'})}\, \varepsilon^{3c-\nu_\psi+\nu_{\psi'}}\ll 1.
\eea

We note that it is also possible to get a hierarchical Yukawa coupling without a large warp factor, as far as the bulk mass parameters, $\nu_\psi, \nu_{\psi'}$, are parametrically larger than unity, namely, the condition $\varepsilon^{3c-\nu_\psi+\nu_{\psi'}}\ll 1$ would be sufficient.

\section{Bulk Gauge Bosons}
 
 We consider a massless bulk gauge boson for either abelian or non-abelian gauge symmetry in the linear dilaton background in five dimensions \cite{CWphoton}.
 A gauge fixing term is required for a massless gauge boson. In the following, we set $A_5=0$, which is suitable for components of the gauge boson containing a massless mode on orbifold $S^1/Z_2$.

The Lagrangian for a bulk gauge boson $A_M$ in Jordan frame is given by
\bea
{\cal L}_A=- \sqrt{G}\, e^{a S}\,  \frac{1}{4} F_{MN} F_{PQ} G^{MP} G^{NQ}  \label{gauge}
\eea 
where $a$ is a constant parameter for the dilaton coupling, and $F_{MN}=\partial_M A_N -\partial_N A_M$. 
 Then, the Euler equation is
 \bea
 \partial_M \Big( \sqrt{G} \, e^{a S}\, F^{MN}\Big)=0. \label{gauge-eq}
 \eea
 
 On the other hand, in Einstein frame, we can rewrite the bulk Lagrangian (\ref{gauge}) in the following,
 \bea
{\cal L}_A=- \sqrt{G_E}\, e^{(a-\frac{1}{3})S}\, \frac{1}{4} F_{MN} F_{PQ}  G^{MP}_E G^{NQ}_E 
\eea 
Then, the corresponding Euler equation is
\bea
 \partial_M \Big( \sqrt{G_E} \, e^{(a-\frac{1}{3})S}\,  G^{MP}_E G^{NQ}_E F_{PQ} \Big)=0.
 \eea

\subsection{Gauge clockwork modes}

For the flat metric in Jordan frame, $G_{MN}=\eta_{MN}$, and in the gauge with $A_5=\partial_\mu A^\mu=0$ \footnote{We can first choose $A_5=0$ by a 5D gauge transformation with $A_M\rightarrow A_M+\partial_M\alpha$, and perform another gauge transformation satisfying $\partial_5 \alpha'=0$ and $\partial_\mu A^{\prime\mu}=\partial_\mu A^\mu+\Box \alpha'=0$. For massless mode, we can perform one more gauge transformation with $\Box\alpha^{\prime\prime}=0$ to reduce the number of polarization states to two. But,  for massive modes, there remain three polarization states. }, the bulk equation (\ref{gauge-eq}) for the gauge field in Jordan frame becomes
\bea
\Box A_\mu -a S' \partial_z A_\mu - \partial^2_z A_\mu =0. 
\eea
Thus, the above equation is the same as  the one for a massless bulk scalar in the CW model in eq.~(\ref{CWscalar}), up to the dilaton coupling. 
So, making a Fourier decomposition of the bulk gauge field as $A_\mu(x,z)=e^{-ka |z|} \sum_n A^{(n)}_\mu(x) f^A_n(z) $ with $(\Box+m^2_n)A^{(n)}_\mu(x)=0$, 
we obtain the similar eigenfunctions and eigenvalues as those for a massless bulk scalar, as follows,
\bea
f^A_0&=& N_{A_0} \,e^{ka |z|}, \label{gauge-zero} \\
f^A_n&=& N_{A_n} \bigg(\cos\frac{\pi n z}{z_c}+\frac{ka z_c}{\pi n}\,\sin\frac{\pi n|z|}{z_c} \bigg), \quad n\in Z,
 \label{gauge-massive}
\eea
with
\bea
N_{A_0}&=& \sqrt{\frac{ka }{e^{2ka z_c}-1}},  \label{gnorm} \\
 N_{A_n} &=&\frac{1}{\sqrt{z_c}}\, \Big(\frac{\pi n}{z_c m_{A_n}}\Big), \label{gnorm2}
 \eea
 and 
\bea
m^2_{A_n}&=& a^2 k^2 + \frac{\pi^2 n^2}{z^2_c}.
\eea
Therefore, the KK masses of the bulk gauge boson depend on the 5D curvature scale $k$ and the radius of the extra dimension $z_c$ as well as the dilaton coupling $a$.

We note that in the Gaussian normal coordinate $y$, the zero mode solution for the bulk gauge field becomes $f^A_0=N_{A_0}\, (\frac{2}{3} k|y|+1)^{\frac{3}{2}a}$ with $N_{A_0}=\sqrt{ka/[(\frac{1}{3} kL_5 +1)^{3a}-1]}$. Thus, the exponential warp factor is replaced by the power-law factor as in the zero modes for bulk scalar and fermion fields, but with a large proper length of the extra dimension in the normalization factor.

\subsection{Localized couplings of gauge clockwork}

We can introduce a  coupling of the bulk gauge field to the external charged fields, $\chi_i, \psi_i$, localized at $z=z_i (i=1,2)$, with $z_1=0, z_2=z_c$, written in Jordan frame,
\bea
{\cal L}_{A,{\rm brane}} &=& \frac{\sqrt{G}}{\sqrt{-G_{55}}}\, \sum_{i=1,2}\delta(z-z_i)\, J^{\mu,i}(x) A_\mu(x,z).
\eea
Here, we have not introduced the dilaton couplings to the brane-localized charged fields by imposing the universality of the zero-mode gauge couplings independent of the locations in the bulk.
Then, from the Fourier expansion of the bulk gauge field, we obtain the effective gauge couplings in Einstein frame as
\bea
{\cal L}_{A,{\rm brane}} 
&=& J^{\mu,i}(x) e^{-k a z_i}  \Big ( f^A_0(z_i)\, A_{\mu,0}(x) + f^A_n(z_i) A^{(n)}_{\mu}(x) \Big) \nonumber \\
&=& \Big(-i q_\chi\,  \chi_i \partial^\mu \chi^*_i+{\rm h.c.}+q_\psi {\bar\psi}_i \gamma^\mu \psi_i\Big) \Big( g_0 \, A_{\mu,0}(x)+g_n\, A^{(n)}_{\mu}(x)  \Big)
\eea
where the brane charged current is given by $J^{\mu,i} =-i q_\chi g_{5D} \chi_i \partial^\mu \chi^*_i+{\rm h.c.} +q_\psi g_{5D} {\bar\psi}_i \gamma^\mu \psi_i$, and the mode functions of the bulk gauge field are given in eqs.~(\ref{gauge-zero}) and (\ref{gauge-massive}).
Then, the 4D effective gauge coupling for the zero mode of the bulk gauge boson is given by
\bea
g_0= g_{5D} \sqrt{\frac{ka}{e^{2kaz_c}-1}}, \label{zero-gauge}
\eea
and the effective couplings for the massive modes of the bulk gauge field depend on the locations of the charged fields, given by
\bea
g_n=g_{5D} \frac{1}{\sqrt{z_c}}\, \frac{\pi n}{z_c m_{A_n}}\times \left\{\begin{array}{cc} 1, \quad  z=0, \\ (-1)^n\, e^{-k az_c}, \quad z=z_c. \end{array}\right. \label{brane-gmassive}
\eea
Therefore, the effective couplings of massive modes of the bulk gauge boson to the external charged fields at $z=z_c$ are exponentially suppressed, relative to those at $z=0$.

We remark on the dependences of the effective gauge couplings on the dilaton coupling of the bulk gauge field $a$ and the warp factor.
First, for $a>0$ and $e^{k|a|z_c}\gg 1$ as required by the solution to the hierarchy problem, the zero-mode gauge coupling becomes $g_0\simeq g_{5D} \sqrt{k|a|} \, e^{-k|a|z_c}$,  from eq.~(\ref{zero-gauge}), which is too suppressed to be the observed value of the gauge coupling in the SM, unless $g_{5D}$ is taken to a large value.

Second, for $a<0$ and $e^{k|a|z_c}\gg 1$, the zero-mode gauge coupling becomes $g_0\simeq g_{5D} \sqrt{k|a|} $,  from eq.~(\ref{zero-gauge}), which can be chosen to the observed value of the gauge coupling  in the SM, without a need of taking a large value of $g_{5D}$. 
However, in this case, the localized charged particles at $z=z_c$ have enhanced effective gauge couplings for the massive modes of the bulk gauge field in eq.~(\ref{brane-gmassive}), due to the exponential factor, $e^{k|a|z_c}$. Then, the result is questionable for perturbativity. Therefore, in order to maintain perturbativity being compatible with the hierarchy problem, we would need to introduce the charged fields for the bulk gauge field on the brane  at $z=0$. 
Otherwise, we need to take $k|a| z_c\sim 1$ for $|a|\lesssim 0.1$. 
But, if $|a|$ is sizable, we need to take $k z_c={\cal O}(1)$ for perturbativity throughout the bulk. 

Lastly, for $a=0$, which means that there is no dilaton coupling to the bulk gauge field, we get the zero-mode gauge coupling as $g_0=g_{5D}/\sqrt{2z_c}$ from eq.~(\ref{zero-gauge}) as in the flat extra dimension, and the massive-mode gauge couplings are given by  $|g_n|=\sqrt{2} g_0$ from eq.~(\ref{brane-gmassive}), independent of the branes, and the KK masses become $m_{A_n}=\pi n/z_c$.
In this case, the masses of KK gauge bosons would be lighter than those of bulk gravitons or fermions at least by the order of magnitude for $kz_c={\cal O}(10)$.

In the next subsection, we will give a more general discussion on the bulk gauge couplings from the localized zero modes of bulk charged scalars and fermions.

\subsection{Bulk couplings of gauge clockwork}

The bulk gauge field also couples to the external scalar or fermion fields, $\chi,\psi$, living in the bulk, whose gauge interactions are written in Jordan frame,
\bea
{\cal L}_{A,{\rm bulk}}= \sqrt{G}\,  e^{c S}\, J^{\mu}(x,z) A_\mu(x,z)
\eea
where the bulk charged current is given by $J^{\mu}(x,z)=-i q_\chi g_{5D} \chi \partial^\mu \chi^*+{\rm h.c.}+q_\psi g_{5D} {\bar\psi} \gamma^\mu \psi$.
Then, for the zero modes of the external fields,  $\chi_0, \psi_0$, we obtain the following effective gauge interactions to the KK modes of the bulk gauge field in Einstein frame,
\bea
{\cal L}_{A,{\rm eff}}&=& \Big(-i q_\chi  \chi_0\partial_\mu \chi^*_0+{\rm h.c.}\Big) \Big(g_0 A^{(0)}_\mu
 +  g_{\chi,n} A^{(n)}_\mu \Big)   \nonumber \\
&&+ q_\psi {\bar\psi}_0 \gamma^\mu \psi_0 \Big(g_0 A^{(0)}_\mu +  g_{\psi,n} A^{(n)}_\mu \Big)
\eea
where
\bea
 g_{\chi,n} &=&g_{5D} N_{A_n} N^2_{\chi_0} \int^{z_c}_{-z_c} dz\, e^{k(2c-a)|z|}\, \bigg(\cos\frac{\pi n z}{z_c}+\frac{ka z_c}{\pi n}\,\sin\frac{\pi n|z|}{z_c}\bigg) \nonumber \\
 &=& g_{5D}\, \frac{1}{\sqrt{z_c}}\,\frac{\pi n}{z_c m_{A_n}}\,  \frac{4 c(a-c) (kz_c)^2}{|e^{2kc z_c}-1|}\cdot\frac{\Big(1-(-1)^n\, e^{-(a-2c)k z_c}\Big)}{n^2\pi^2+ (a- 2c)^2 (k z_c)^2},  \label{s-massive} \\
  g_{\psi,n} &=& g_{5D} N_{A_n} N^2_{L,R}  \int^{z_c}_{-z_c}  dz\, e^{-ka|z|}\, e^{-\frac{2}{3}(-3c\pm 2\nu)k|z|}\, \bigg(\cos\frac{\pi n z}{z_c}+\frac{ka z_c}{\pi n}\,\sin\frac{\pi n|z|}{z_c}\bigg) \nonumber \\
&=&g_{5D}\, \frac{1}{\sqrt{z_c}}\,\frac{\pi n}{z_c m_{A_n}}\,  \frac{4 |-3c\pm 2\nu| (3a-3c\pm 2\nu)(kz_c)^2}{|1-e^{-\frac{2}{3}(-3c\pm 2\nu)kz_c}|}  \nonumber \\
&&\quad\times \frac{\Big(1-(-1)^n\, e^{-\frac{1}{3}(3a+2(-3c\pm 2\nu))k z_c}\Big)}{9n^2\pi^2+ (3a-6c\pm 4\nu)^2 (k z_c)^2}. \label{f-massive}
\eea

We now discuss the impacts of the dilaton couplings, the warp factor and the bulk mass parameter $\nu$ on the obtained effective gauge interactions to the KK modes of the bulk gauge field.
First, taking $a<0$ for a sizable gauge coupling for the zero-mode gauge boson in eq.~(\ref{zero-gauge}) and  $c<\frac{1}{2}a$ for perturbativity, the charged scalar couplings to the massive-mode gauge bosons in eq.~(\ref{s-massive}) become
for $e^{-k z_c}\ll 1$,
\bea
 g_{\chi,n} \approx g_{5D}\, \frac{1}{\sqrt{z_c}}\,\frac{\pi n}{z_c m_{A_n}}\, \frac{4 c(a-c) (kz_c)^2}{n^2\pi^2+ (a- 2c)^2 (k z_c)^2}.
\eea
Thus, in this case, the massive-mode gauge bosons have mildly suppressed couplings by the factor of $(k z_c)^{-3/2}$ as compared to the one for the zero-mode gauge boson, which is approximated to $g_0\approx g_{5D}\sqrt{k|a|}$.

Imposing $a<0$, and $|\nu|<\frac{3}{2}c$ for $c>0$ (or  $|\nu|>\frac{3}{2}|c|$ for $c<0$), the effective couplings of the bulk charged fermion to the massive-mode gauge bosons in eq.~(\ref{f-massive}) become
\bea
  g_{\psi,n}\approx (-1)^{n+1} \,g_{5D}\, \frac{1}{\sqrt{z_c}}\,\frac{\pi n}{z_c m_{A_n}}\,\frac{4 |-3c\pm 2\nu| (3a-3c\pm 2\nu) (kz_c)^2}{9n^2\pi^2+ (3a-6c\pm 4\nu)^2 (k z_c)^2}\,\cdot e^{k|a|z_c}. \label{delocal}
\eea
Consequently, in this case, the charged fermion localized towards $z=z_c$ would have exponentially enhanced couplings for $e^{k|a|z_c}\gg1$, unless $|a|$ is small. Thus, the warp factor would be bounded to $kz_c={\cal O}(1)$ by perturbativity or we would need to take a small $|a|$ such that $|a|\lesssim 1/(kz_c)$.

Finally, taking $a<0$ and $|\nu|>\frac{3}{2}c$ for $c>0$  instead (or $|\nu|<\frac{3}{2}|c|$ for $c<0$) with  $3a+2(-3c\pm 2\nu)>0$, we can approximate the effective couplings of the bulk charged fermion to the massive-mode gauge bosons in eq.~(\ref{f-massive}) as
\bea
  g_{\psi,n}\approx g_{5D}\, \frac{1}{\sqrt{z_c}}\,\frac{\pi n}{z_c m_{A_n}}\,\frac{4 |-3c\pm 2\nu|(3a-3c\pm 2\nu) (kz_c)^2}{9n^2\pi^2+ (3a-6c\pm 4\nu)^2 (k z_c)^2}, \label{local}
\eea
which is mildly suppressed by the factor of $(k z_c)^{-3/2}$ as compared to  the zero mode coupling, $g_0\simeq g_{5D} \sqrt{k|a|}$.
Thus, in this case, we can take a large warp factor as required for solving the hierarchy problem without a problem of perturbativity.

\section{Bulk Gravitons}

We consider the bulk graviton ${\hat G}_{\mu\nu}(z,x)$ as the perturbation around the warped metric in the Einstein frame, 
\bea
ds^2=w(z)^2 \Big[(\eta_{\mu\nu}+{\hat G}_{\mu\nu}(z,x) ) dx^\mu dx^\nu + dz^2\Big].
\eea
Under the Fourier decomposition of the 5D metric as 
\be
{\hat G}_{\mu\nu}(z,x)= 2M^{-3/2}_5 w^{-2} f^G_n(z) G^{(n)}_{\mu\nu}(x)
\ee
 with $(\Box-m^2_{G_n})G^{(n)}_{\mu\nu}(x)=0$, the bulk linearized Einstein equation leads to the equation for the mode functions of the bulk graviton in the extra dimension \cite{5d-mpert}, 
\bea
(f^{G}_n)^{\prime\prime}-\frac{w'}{w}\,(f^{G}_n)^{\prime} +\Big(m^2_{G_n}-\frac{2w^{\prime\prime}}{w} \Big)f^G_n=0. \label{graviton}
\eea

\subsection{Graviton clockwork modes}

For the warp factor, in the CW model $w(z)=e^{\frac{2}{3}k|z|}$, we obtain the equation for the mode functions from eq.~(\ref{graviton}) as
\bea
(f^G_n)^{\prime\prime}-\frac{2}{3} k (f^G_n)^\prime +\Big(m^2_{G_n}-\frac{8}{9}k^2 \Big)f^G_n-\frac{8}{3}k\,(\delta(z)-\delta(z-z_c)) f^G_n=0. 
\eea
Then, making the field redefinition with $f^G_n=e^{\frac{1}{3}k|z|}\psi_n$, we get the above equation as
\bea
\psi^{\prime\prime}_n +(m^2_{G_n} -k^2) \psi_n -2k(\delta(z)-\delta(z-z_c))\psi_n=0. 
\eea
For the range of the extra dimension to be $z\in[-z_c,z_c]$,  the boundary conditions for the mode functions are given by
\bea
\Big(\psi^\prime_n - k\psi_n\Big)\Big|_{z=0^+}&=&0, \label{bc3} \\
\Big(\psi^\prime_n - k\psi_n \Big)\Big|_{z=z^-_c}&=&0. \label{bc4}
\eea
As a result, we obtain the solution for the massless mode with $m_0=0$ satisfying the boundary conditions, (\ref{bc3}) and (\ref{bc4}) as
\bea
\psi_0(z)= C_0\, e^{k|z|}
\eea
with 
\bea
C_0=\sqrt{\frac{k}{e^{2k z_c}-1}}. 
\eea
On the other hand, the solutions to massive modes with $m_n\neq 0$ are
\bea
\psi_n(z)= C_n\Big(\cos \frac{\pi nz}{z_c}+\frac{kz_c}{\pi n}\sin\frac{\pi n|z|}{z_c} \Big), \quad n\in Z,
\eea
with 
\bea
C_n &=&\frac{1}{\sqrt{z_c}}\, \frac{\pi n}{m_{G_n} z_c}, \\
m^2_{G_n}&=&k^2+\frac{\pi^2 n^2}{z^2_c}. 
\eea
Therefore, the KK masses of the bulk graviton has a mass gap determined by the 5D curvature scale $k$.

\subsection{Localized couplings of graviton clockwork}

We can introduce a  coupling of the bulk graviton to the external fields, $\chi_i, \psi_i$, localized at $z=z_i (i=1,2)$, with $z_1=0, z_2=z_c$, written in terms of the energy-momentum tensors in Einstein frame,
\bea
{\cal L}_{G,{\rm brane}} &=& -  \frac{1}{2} \frac{\sqrt{G_E}}{\sqrt{-G^E_{55}}} \sum_{i=1,2}\, T^{\mu\nu,i}(x) \Big(\omega^2(z_i) {\hat G}_{\mu\nu}(x,z_i)\Big) \nonumber \\
&=&- \sum_{i=1,2}\,T^{\mu\nu,i}(x) \Big (\frac{1}{M_P}\, G^{(0)}_{\mu\nu}(x) +  \frac{1}{\Lambda^i_n}\,G^{(n)}_{\mu\nu}(x) \Big)
\eea
where $ T^{\mu\nu,i}(x)$ are the energy-momentum tensors for the external fields localized at $z=z_i$, and the KK graviton couplings \cite{CWgraviton,CWg2} are given by
\bea
\frac{1}{\Lambda^i_n}&=&M^{-3/2}_5e^{-kz_i}\, \frac{1}{\sqrt{z_c}}\,\frac{\pi n}{m_{G_n} z_c} \cos\Big(\frac{\pi n z_i}{z_c}\Big) \nonumber \\
&\equiv& \frac{e^{-kz_i}}{\Lambda}\,\frac{n m_{G_1}}{m_{G_n}} \cos\Big(\frac{\pi n z_i}{z_c}\Big),  \label{graviton-brane}
\eea
normalized to the suppression scale for the first KK graviton at $z=0$,
\bea
\Lambda\equiv (M_5 z_c)^{3/2}\, \frac{m_{G_1}}{\pi}.
\eea
Thus, the KK gravitons couple strongly to the external field localized at $z=0$, but their couplings to those localized at $z=z_c$ are exponentially suppressed by the order of the inverse Planck scale from $M^{-3/2}_5 e^{-kz_c}/\sqrt{z_c}\sim 1/M_P/\sqrt{kz_c}$, which was obtained from eq.~(\ref{Pmass-CW}). 

In view of the condition for solving the hierarchy problem with $M_5\sim 10\,{\rm TeV}$  and $e^{-2kz_c}\ll 1$ in eq.~(\ref{hierarchy}), we can infer the suppression scale for the KK graviton couplings as
\bea
\frac{\Lambda}{M_5} \approx 182\, \Big(\frac{M_5}{10\,{\rm TeV}}\Big)^{1/2} \Big(\frac{k z_c}{32} \Big)^{3/2} \Big(\frac{1\,{\rm TeV}}{k}\Big)^{1/2}.
\eea

\subsection{Bulk couplings of graviton clockwork}

The bulk graviton also couples to the external fields in the bulk, written in terms of the energy-momentum tensors in Einstein frame,
\bea
{\cal L}_{G,{\rm bulk}}=-    \frac{1}{2}\, \sqrt{G_E}\, T^{\mu\nu}(x,z) \Big(\omega^2(z) {\hat G}_{\mu\nu}(x,z)\Big).
\eea
Then, for the zero modes of the external fields, we obtain the following effective  interactions to the KK gravitons,
\bea
{\cal L}_{G,{\rm eff}}= -T^{\mu\nu}_{(0)}(x) \bigg(\frac{1}{M_P}\,G^{(0)}_{\mu\nu}(x)+\frac{1}{\Lambda^B_n}\, G^{(n)}_{\mu\nu}(x)  \bigg):
\eea
for scalar fields,
\bea
\frac{1}{\Lambda^B_n}=M^{-3/2}_5\,N^2_{\chi_0} \int^{z_c}_{-z_c} dz\,  e^{(2c-1)k|z|} \psi_n(z) \equiv \frac{c_{\chi,n}}{\Lambda}
 \label{graviton-bulk}
\eea
with
\bea
c_{\chi,n}&=&\frac{n m_{G_1}}{m_{G_n}}\, \frac{4|c|(1-c)(k z_c)^2}{|e^{2k c z_c}-1|} \cdot \frac{\Big(1-(-1)^n\, e^{(2c-1)k z_c}\Big)}{n^2\pi^2+ (1-2c)^2 (k z_c)^2}\,;
\eea
for gauge bosons,
\bea
\frac{1}{\Lambda^B_n}=M^{-3/2}_5\,N^2_{A_0} \int^{z_c}_{-z_c} dz\,  e^{(2a-1)k|z|} \psi_n(z) \equiv \frac{c_{A,n}}{\Lambda}
 \label{graviton-bulk2}
\eea
with
\bea
c_{A,n}&=&\frac{n m_{G_1}}{m_{G_n}}\, \frac{4|a|(1-a)(k z_c)^2}{|e^{2k a z_c}-1|} \cdot \frac{\Big(1-(-1)^n\, e^{(2a-1)k z_c}\Big)}{n^2\pi^2+ (1-2a)^2 (k z_c)^2}\,;
\eea
for fermions,
\bea
\frac{1}{\Lambda^B_n}=M^{-3/2}_5\, N^2_{L,R}\int^{z_c}_{-z_c} dz\,  e^{-k|z|}  e^{-\frac{2}{3}(-3c\pm 2\nu)k|z|}\, \psi_n(z) 
\equiv \frac{c_{\psi,n}}{\Lambda}
  \label{graviton-bulk3}
\eea
with
\bea
c_{\psi,n}
=\frac{n m_{G_1}}{m_{G_n}}\,\frac{4|-3c\pm 2\nu|(3-3c\pm 2\nu) (kz_c)^2}{\big|1-e^{-\frac{2}{3}(-3c\pm 2\nu)kz_c}\big|}\, \cdot \frac{\Big(1-(-1)^n\, e^{-\frac{1}{3}(3+2(-3c\pm 2\nu))k z_c}\Big)}{9n^2\pi^2+ (3-6c\pm 4\nu)^2(k z_c)^2}.
\eea
Here, $N_{\chi_0}, N_{A_0}, N_{L,R}$ are the normalization factors for zero modes, given in eqs.~(\ref{bnorm}), (\ref{gnorm}), (\ref{fnorm}), respectively.

As a result, first, from eq.~(\ref{graviton-bulk}) that the KK graviton couplings to the zero mode of the bulk scalar are exponentially suppressed for $c>0$, but they are comparable to those to the fields localized at $z=0$  for $c<0$, that is, $\Lambda^1_n$ in eq.~(\ref{graviton-brane}).
We note that for $c=0$ or $c=1$, the latter of which is the same as the one for the dilaton field in eq.~(\ref{jordan}),  the KK graviton couplings vanish identically. Second, from eq.~(\ref{graviton-bulk2}), the KK graviton couplings to the zero mode of the bulk gauge boson has a similar dependence on the dilaton coupling $a$ as for the bulk scalar field.

Finally, from eq.~(\ref{graviton-bulk3}), the KK graviton couplings to the zero mode of the bulk fermion are comparable to those to the fields localized at $z=0$, that is, $\Lambda^1_n$ in eq.~(\ref{graviton-brane}), for $|\nu|>\frac{3}{2}c$ with $c>0$ (which corresponds to the localization towards $z=0$), whereas being exponentially suppressed, similarly to $\Lambda^2_n$ in eq.~(\ref{graviton-brane}),  for $|\nu|<\frac{3}{2}c$ with $c>0$ (which corresponds to the localization towards $z=z_c$).

\section{The Clockwork Standard Model}

In this section, we make use of the results for the mode functions and couplings of the bulk fields in the linear dilaton background in the previous sections and construct the bulk SM Lagrangian, which is regarded as the continuum limit of the Clockwork SM.

We assume that the electroweak symmetry is broken due to the VEV of the Higgs doublet localized on the brane at $z=0$. Then, in order to explain the mass hierarchy of the SM fermions, we also assume that the SM fermions, in particular, the light fermions other than top quark and/or bottom quark, propagate into the bulk such that the effective Yukawa couplings for them are suppressed.

The full Lagrangian for the Clockwork SM including the right-handed neutrinos $n_R$ are then given  in Jordan frame by
\bea
{\cal L}_{\rm CW\,SM} &=& \sqrt{G} \,  e^{c S}\bigg[\sum_{\psi=q,u,d,l,e, n}i{\bar \psi} \Gamma^M\Big( D_M +\frac{1}{8} \omega_M\,^{\underline{A}\underline{B}}[\Gamma_{\underline{A}},\Gamma_{\underline{B}}]  \Big)\psi -e^{\frac{1}{3}S}\,m_{\psi}{\bar \psi} \psi\bigg] \nonumber \\
&&- \sqrt{G} \,  e^{a S}\bigg[\frac{1}{4}B_{\mu\nu} B^{\mu\nu} +\frac{1}{2} {\rm Tr}(W_{\mu\nu} W^{\mu\nu}) +\frac{1}{2} {\rm Tr} (g_{\mu\nu} g^{\mu\nu}) \bigg] \nonumber \\
&&+ \delta(z)\, \frac{ \sqrt{G} }{\sqrt{-G_{55}}}\, \Big(|D_\mu H|^2- V(H)  -y_d\,  {\bar q}_L d_R H-y_u\,  {\bar q}_L u_R {\widetilde H} -y_e\, {\bar l}_L e_R H +{\rm h.c.}\nonumber \\
&&\quad - y_\nu\, {\bar l}_L n_R {\widetilde H} - \frac{1}{2} M_R \overline{n^c_R} n_R +{\rm h.c.} \Big). \label{bulk}
\eea
where  the Higgs potential is given by $V(H)=m^2_H |H|^2 +\lambda_H |H|^4$, the 5D covariant derivatives for the SM fermions are given by $D_M=\partial_M-i g_{Y,5D}  Y B_\mu(x,z)- i g_{L,5D} W_\mu(x,z)-ig_{S,5D} g_\mu(x,z)$ with $W_\mu=\frac{1}{2} {\vec \tau}\cdot {\vec W}_\mu, g_\mu=\frac{1}{2}\lambda^a g^a_\mu $ and $g_{Y,5D}, g_{L,5D}, g_{S,5D}$ being the bulk gauge couplings for $U(1)_Y$,  $SU(2)_L$, $SU(3)_C$, respectively, and the mass parameters for bulk fermions with $m_{\psi}=\nu_{\psi} \sigma'$ can be independently chosen, and ${\widetilde H}=i\tau^2 H^*$.  
We note that $\lambda^a$ are Gell-Mann matrices and $\tau^i$ are Pauli matrices, satisfying $[\frac{\lambda^a}{2},\frac{\lambda^b}{2}]=if^{abc} \frac{\lambda^c}{2} $, $[\frac{\tau^i}{2},\frac{\tau^i}{2}]=i\epsilon^{ijk} \frac{\tau^k}{2}$, as well as ${\rm Tr}(\lambda^a \lambda^b)=2 \delta^{ab}$ and ${\rm Tr}(\tau^i \tau^j)=2 \delta^{ij}$.
We also note that the dilaton couplings to the bulk fermions and gauge bosons are introduced universally as $c$ and $a$, respectively.

Some of the matter fermions and/or gauge fields in the SM can be localized on the branes.
In this case, we can consider the brane-localized kinetic terms for them in the following form,
\bea
{\cal L}_b= \delta(z-z_i)\,\frac{\sqrt{G}}{\sqrt{-G_{55}}} \,\bigg(i{\bar\psi}_i \gamma^\mu D_\mu\psi_i - \frac{1}{4} \, G^{\mu\rho} G^{\nu\sigma} F_{i,\mu\rho} F_{i,\nu\sigma}  \bigg).
\eea
In this case, for the matter fermions localized on the branes, we don't include the normalization factors for them in writing the effective Yukawa couplings, unlike those for the zero modes of bulk fermions.

\begin{figure}[tbp]
\centering 
\includegraphics[width=.45\textwidth]{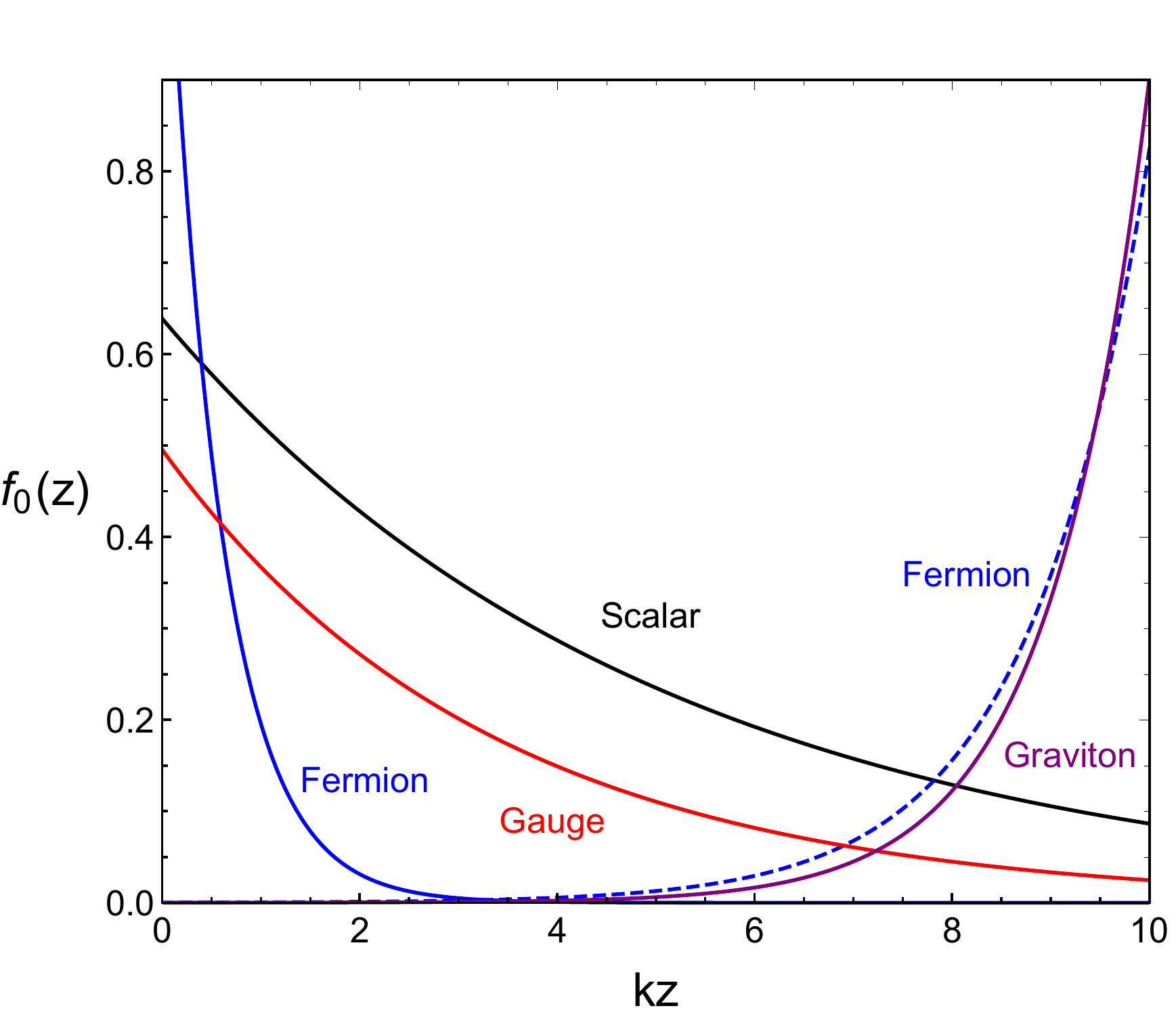} \,\,\,\,
\includegraphics[width=.45\textwidth]{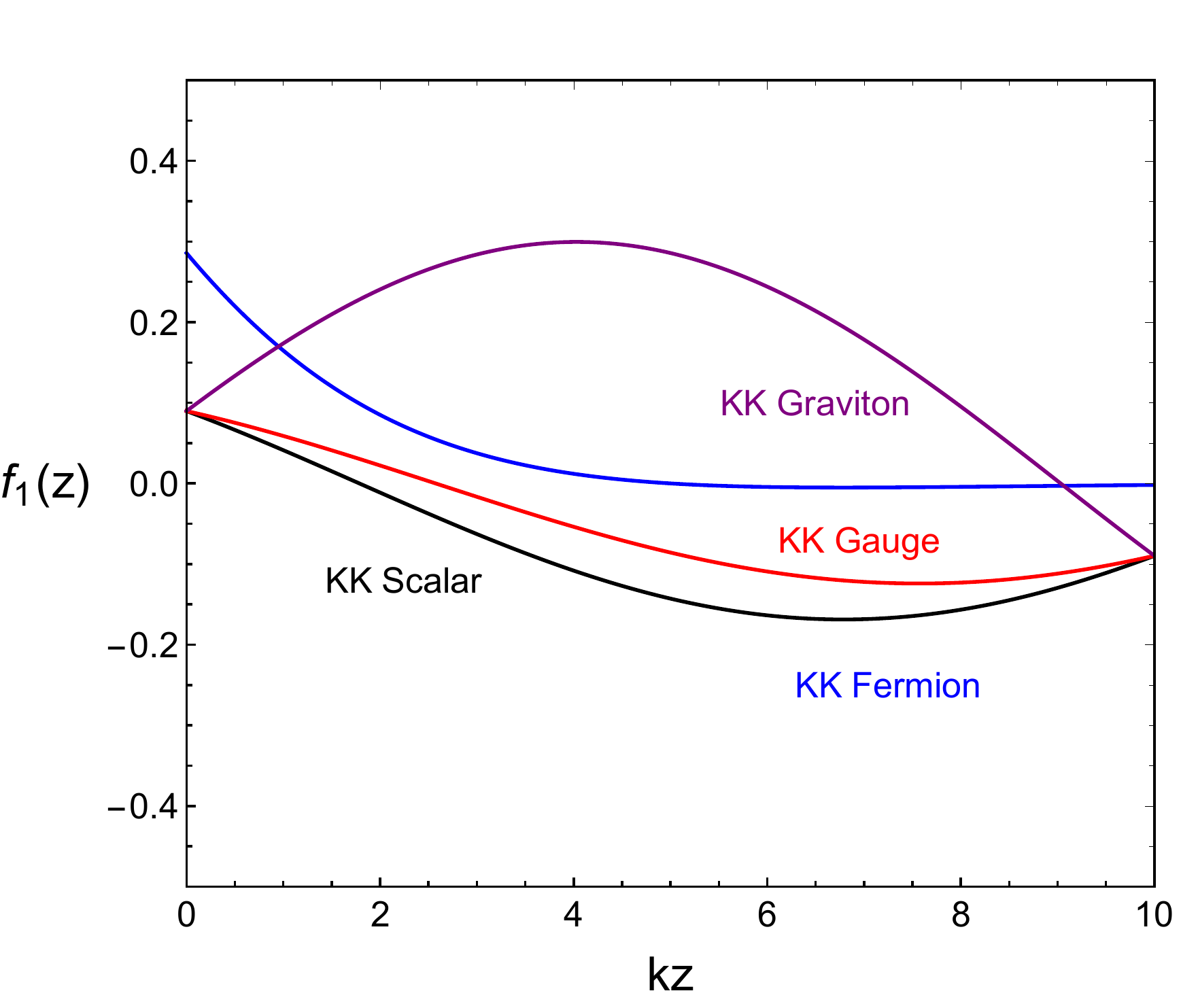} 

\caption{\label{zeromode} (Left) The bulk profiles of zero modes of bulk fields. (Right) The bulk profiles of the first KK modes of bulk fields. The wave functions for scalar, left-handed fermion, gauge boson and graviton are shown in black, (dashed) blue, red and purple lines, respectively. We took the bulk mass parameters, $\nu=\pm 2$, for solid and dashed blue lines, and the dilaton couplings, $c=-0.5$ for scalar and fermion, and $a=-0.3$ for gauge boson, and $kz_c=10$. }
\end{figure}

In Fig.~\ref{zeromode}, we show the schematic bulk profiles of  the zero modes and the first KK modes for scalar, left-handed fermion, gauge boson as well as graviton, with arbitrary common normalization. 
We have taken the bulk mass parameters for the bulk fermion to $\nu=\pm 2$ for solid and dashed blue lines, and the dilaton couplings are chosen to $c=-0.5$ for scalar and fermion and $a=-0.3$ for gauge boson, and the parameter for the warp factor is  $kz_c=10$, as illustration. In this case, the zero modes of scalar, gauge boson and left-handed fermion with $\nu=2$ (corresponding to heavy fermions) are localized towards $z=0$, whereas the zero modes of graviton and left-handed fermion with $\nu=-2$ (corresponding to light fermions) are localized towards $z=z_c$.
On the other hand, the first KK modes for all the cases are distributed through the bulk.

\subsection{The Yukawa couplings for quarks and leptons}

For a bulk fermion with the mass parameter $\nu_\psi$, from eqs.~(\ref{fermion1}) and (\ref{fermion2}) with eqs.~(\ref{fzero}) and (\ref{fnorm}), the zero modes as the probability densities with $2\int^{z_c}_0 dz\, (e^{-\frac{3}{2}c\sigma}\psi_0)^2=1$  in the $z$ coordinate are given by
\bea
e^{-\frac{3}{2}c\sigma}\psi_0(x,z)&=& \left\{\begin{array}{c}  N_{\psi_L}\,  \psi_{L,0}(x)\, e^{-\frac{1}{3}(-3c+2\nu_\psi)k |z|},\qquad\quad\quad \psi=q,l, \vspace{0.2cm}\\
N_{\psi_R}\, \psi_{R,0}(x)\, e^{-\frac{1}{3}(-3c-2\nu_\psi)k |z|} ,\qquad \psi=u,d, e, n \end{array}\right.
\eea
where
\bea
N_{\psi_L} &=&\sqrt{\frac{\frac{1}{3}(-3c+2\nu_\psi) k}{ 1- e^{-\frac{2}{3}(-3c+2 \nu_\psi) k z_c}}}, \\
N_{\psi_R} &=&\sqrt{\frac{\frac{1}{3}(-3c-2\nu_\psi) k}{ 1- e^{-\frac{2}{3}(-3c-2 \nu_\psi) k z_c}}}.
\eea
Then, for the mass hierarchy of fermions, the light fermions must be delocalized from the brane at $z=0$, so  we need to choose $\nu_\psi<\frac{3}{2}c$ for $\psi=q, l$, and $\nu_\psi>-\frac{3}{2}c$ for $\psi=u,d,e, n$. On the other hand, the top quark must be localized on the brane at $z=0$, so we need $\nu_{t_L}>\frac{3}{2}c$ and/or $\nu_{t_R}<-\frac{3}{2}c$.
As a result, for $e^{-\frac{2}{3}(-3c\pm 2\nu_\psi)kz_c}\gg 1$, except the top quark, we can approximate $N_{\psi_L}\approx \sqrt{\frac{k}{3}(3c-2\nu_\psi)}\, e^{\frac{1}{3}(-3c+2 \nu_\psi) k z_c}$ and $N_{\psi_R}\approx \sqrt{\frac{k}{3}(3c+2 \nu_\psi)}\, e^{\frac{1}{3}(-3c-2 \nu_\psi) k z_c}$.

As a consequence, after inserting the zero mode wave functions for the bulk fermions in the Yukawa couplings in eq.~(\ref{bulk}), we derive the effective Yukawa couplings as follows,
\bea
-{\cal L}_{\rm Y}&=&y_d\,N_{q} N_{d}\,  {\bar q}_{L,0} d_{R,0} H+y_u\,N_{q} N_{u} \,  {\bar q}_{L,0} u_{R,0} {\widetilde H} +y_e\,N_{l} N_{e} \, {\bar l}_{L,0} e_{R,0} H +{\rm h.c.}  \nonumber \\
&&+ y_\nu\, N_{l} N_{n}\, {\bar l}_L n_R {\widetilde H} + \frac{1}{2} N^2_{n}\, M_R \overline{n^c_R} n_R +{\rm h.c.}  \nonumber \\
&=&\lambda_{ d}\, {\bar q}_{L,0} d_{R,0} H+ \lambda_{u}\, {\bar q}_{L,0} u_{R,0} {\widetilde H} +\lambda_{e}\,  {\bar l}_{L,0} e_{R,0} H +{\rm h.c.}  \nonumber \\
&& + \lambda_{\nu}\, {\bar l}_{L,0} n_{R,0} {\widetilde H} + \frac{1}{2} M'_{R} \overline{n^c_{R,0}} n_{R,0} +{\rm h.c.}  +\cdots
\label{Lfzero}
\eea
with
\bea
\lambda^{ij}_{d} &=&y^{ij}_d\,N^i_{q} N^j_{d}\approx k\,y^{ij}_d  \eta^i_{q} \eta^j_{d},   \\
\lambda^{ij}_{u} &=&y^{ij}_u\,N^i_{q} N^j_{u} \approx k\, y^{ij}_u  \eta^i_{q} \eta^j_{u},   \\
\lambda^{ij}_{e} &=& y^{ij}_e\,N^i_{l} N^j_{e} \approx k\, y^{ij}_e  \eta^i_{l} \eta^j_{e},   \\
\lambda^{ij}_{\nu}&=&  y^{ij}_\nu\,N^i_{l} N^j_{n} \approx  k\, y^{ij}_\nu  \eta^i_{l} \eta^j_{n},   \\
M'_{R,ij} &=& N^i_{n}N^j_{n}\, M_{R,ij} \approx  k M_{R,ij}\, \eta^i_{n} \eta^j_{n}.
\eea
Here, we define the small parameters as
\bea
\eta^i_{q} &=& \sqrt{\frac{1}{3}(3c-2\nu_q)}\, \varepsilon^{\frac{3}{2}c-\nu_q}, \\
\eta^i_{d} &=& \sqrt{\frac{1}{3}(3c+2\nu_{d})}\, \varepsilon^{\frac{3}{2}c+\nu_{d}}, \\
\eta^i_{u} &=& \sqrt{\frac{1}{3}(3c+2\nu_{u})}\,  \varepsilon^{\frac{3}{2}c+\nu_{u}}, \\
\eta^i_{l} &=& \sqrt{\frac{1}{3}(3c-2\nu_{l})}\, \varepsilon^{\frac{3}{2}c-\nu_{l}}, \\
\eta^i_{e} &=& \sqrt{\frac{1}{3}(3c+2\nu_{e})}\, \varepsilon^{\frac{3}{2}c+\nu_{e}}, \\
\eta^i_{n} &=& \sqrt{\frac{1}{3}(3c+2\nu_{n})}\,  \varepsilon^{\frac{3}{2}c+\nu_{n}}
\eea 
where $\varepsilon\equiv e^{-\frac{2}{3}k z_c}$.
Here,  the expansion parameter $\varepsilon\equiv e^{-\frac{2}{3}k z_c}$ is proportional to the inverse of the proper length $L_5$ by  $\varepsilon\approx (\frac{1}{3} k L_5)^{-1}$, so the small expansion parameter is attributed to a large proper length of the extra dimension. 
Thus, for $\eta^i_\psi\ll 1$, we need $|\nu_\psi|<\frac{3}{2}c$ for left-handed fermions for $c>0$ and $|\nu_\psi|>\frac{3}{2}c$ for right-handed fermions for $c<0$.
Then, from eq.~(\ref{kkfmass}), the KK masses for bulk fermions are bounded by $m^2_{\psi_n}=\frac{4}{9}\nu^2k^2+ \frac{\pi^2 n^2}{z_c^2}<c^2k^2+ \frac{\pi^2 n^2}{z_c^2}$ for left-handed fermions for $c>0$ or $m^2_{\psi_n}=\frac{4}{9}\nu^2k^2+ \frac{\pi^2 n^2}{z_c^2}>c^2k^2+ \frac{\pi^2 n^2}{z_c^2}$ for right-handed fermions for $c<0$.

We note that the Yukawa couplings in the bulk SM Lagrangian have an inverse mass dimension, due to the fact that the bulk fermions have a mass dimension two, so the effective Yukawa couplings, $\lambda_{d}, \lambda_{u}, \lambda_{e}, \lambda_{\nu}$, are dimensionless.  On the other hand, the brane Majorana mass $M_R$ is dimensionless, but the effective Majorana mass, $M'_R$, is dimensionful.
We also note that the bulk mass parameters for fermions can be generation dependent, for instance, $m_{q_i}=\nu_{q_i} \sigma'$ with $i=1,2,3$. Then, we can have nontrivial Yukawa matrices as will be discussed later.

As a consequence, even with comparable bulk mass parameters for bulk fermions, we can explain the mass hierarchy and mixing of quarks and leptons, due to the exponential factors. 
Moreover, in the case with lepton number conservation, setting $M_R$ to zero, we can explain the smallness of neutrino masses  for $\nu_{n}+\frac{3}{2}c\gg 1$ due to the exponential suppression of the neutrino Yukawa couplings, thus requiring parametrically larger bulk mass parameters than those for quarks and leptons.

\begin{figure}[tbp]
\centering 
\includegraphics[width=.50\textwidth]{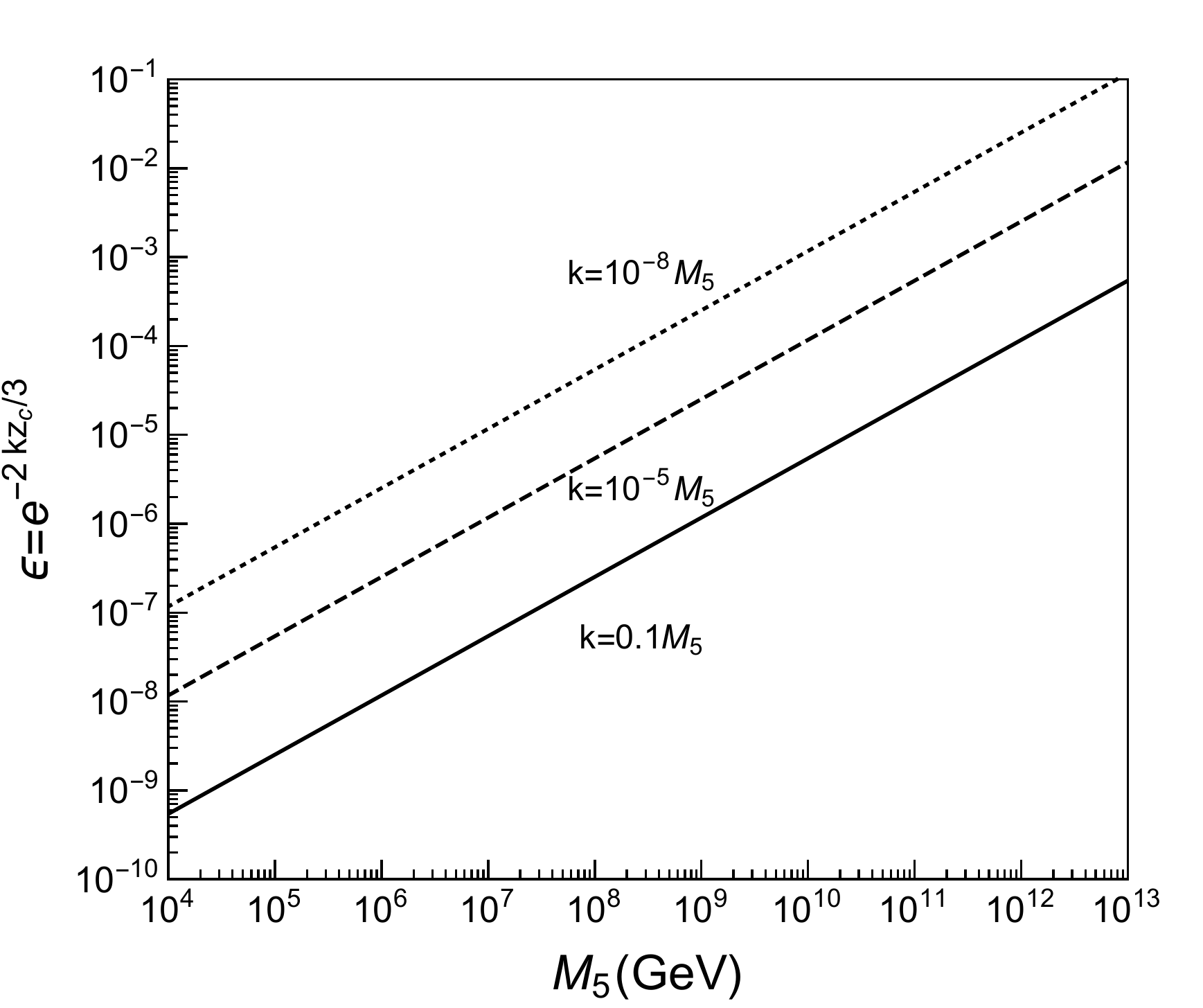} \,\,

\caption{\label{epsilon} The expansion parameter, $\varepsilon\equiv e^{-\frac{2}{3}k z_c}$, as a  function of the 5D Planck mass $M_5$. The 5D curvature scale is chosen to $k=0.1M_5, 10^{-5}M_5, 10^{-8} M_5$, in solid, dashed and dotted lines.  }
\end{figure}

In Fig.~\ref{epsilon}, we depict the small expansion parameter, $\varepsilon\equiv e^{-\frac{2}{3}k z_c}$, as a function of the 5D Planck mass $M_5$.
We have taken the 5D curvature scale to $k=0.1M_5, 10^{-5}M_5, 10^{-8} M_5$, in solid, dashed and dotted lines, in order.
For a small $M_5\sim 10\,{\rm TeV}$, the expansion parameter $\varepsilon$ becomes as small as $\epsilon\sim 10^{-9}$, which would be suitable for explaining the small neutrino masses, as will be discussed in the later subsection.
On the other hand, for a large $M_5$, the expansion parameter $\varepsilon$ can be as large as $\varepsilon\sim 0.1$, which is appropriate for explaining the hierarchy of quark masses and mixings.

\subsection{The mass hierarchy and mixing for quarks}

In this subsection, we discuss the generation of the mass hierarchy and mixing for quarks in the presence of the localizations.

Assuming that the brane-localized Yukawa couplings, $y_d$  and $y_u$, are flavor-diagonal, that is, $y^{ij}_d=y_d\, \delta_{ij}$ and $y^{ij}_u=y_u\, \delta_{ij}$, we want to generate the realistic flavor structure from the localization of the zero modes of bulk fermions. 
In this case, after electroweak symmetry breaking, from eq.~(\ref{Lfzero}), the mass matrices for up-type quarks and down-type quarks are given, respectively, by
\bea
M^{ij}_u &=& \frac{1}{\sqrt{2}}\, k y_u\,v \, \eta^i_{q} \eta^j_{u}, \\
M^{ij}_d &=& \frac{1}{\sqrt{2}}\, k y_d\,v \, \eta^i_{q} \eta^j_{d}.
\eea
Then, assuming a mild hierarchy with $\eta^1_\psi<\eta^2_\psi<\eta^3_\psi$ for $\psi=q,u,d$, we can diagonalize the quark mass matrices by the bi-unitary transformations \cite{hmlee,rotation}, 
\bea
V_{u_L} M_u V^\dagger_{u_R} &=& {\rm diag}(m_u,m_c,m_t)\equiv m^i_u\, \delta_{ij}, \\
V_{d_L} M_d V^\dagger_{d_R} &=& {\rm diag}(m_d,m_s,m_b)\equiv m^i_d\, \delta_{ij},
\eea
where
\bea
V^{ij}_{u_L} &\sim&  V^{ij}_{d_L}\sim {\rm min}\Big(\frac{\eta^i_q}{\eta^j_q}, \frac{\eta^j_q}{\eta^i_q}\Big)\sim  {\rm min}\Big(\varepsilon^{|\nu^i_q|-|\nu^j_q|},\varepsilon^{|\nu^j_q|-|\nu^i_q|}\Big), \label{qrot1} \\
V^{ij}_{u_R} &\sim& {\rm min}\Big(\frac{\eta^i_u}{\eta^j_u}, \frac{\eta^j_u}{\eta^i_u}\Big)\sim  {\rm min}\Big(\varepsilon^{|\nu^i_u|-|\nu^j_u|},\varepsilon^{|\nu^j_u|-|\nu^i_u|} \Big), \\
V^{ij}_{d_R} &\sim& {\rm min}\Big(\frac{\eta^i_d}{\eta^j_d}, \frac{\eta^j_d}{\eta^i_d}\Big)\sim  {\rm min}\Big(\varepsilon^{|\nu^i_d|-|\nu^j_d|},\varepsilon^{|\nu^j_d|-|\nu^i_d|} \Big),
\eea
and $m^i_u=k y_u\, v \,\eta^i_q\eta^i_u$ and $m^i_d=k y_d\, v\, \eta^i_q\eta^i_u$.
Here, we have ignored the quadratic terms for the ratios, $\eta^i_q/\eta^j_q$, etc, but they can be important for a precise matching to the measured CKM matrix. 

Thus, for $ky_u\sim 1$ and $\eta^3_q\eta^3_u\sim 1$ (that is, $|\nu^3_q|+|\nu^3_u|\sim 1$), we obtain the correct top quark mass.
On the other hand, for $ky_d\sim 1$ and $\eta^3_q\eta^3_d\sim \frac{m_b}{m_t}$, we also get the correct bottom mass. 
Moreover, the mass hierarchies for quarks are given by
\bea
\frac{m^i_u}{m^j_u}&=& \varepsilon^{|\nu^i_q|-|\nu^j_q|}\,\times\,  \varepsilon^{|\nu^i_u|-|\nu^j_u|}, \quad i<j, \label{umass} \\
\frac{m^i_d}{m^j_d}&=& \varepsilon^{|\nu^i_q|-|\nu^j_q|}\,\times\,  \varepsilon^{|\nu^i_d|-|\nu^j_d|}, \quad i<j. \label{dmass}
\eea
From eq.~(\ref{qrot1}), the CKM matrix is also obtained as
\bea
V^{ij}_{\rm CKM}= (V_{u_L} V^\dagger_{d_L} )_{ij}=  \frac{\eta^i_q}{\eta^j_q}\sim \varepsilon^{|\nu^i_q|-|\nu^j_q|}, \quad  i<j, \label{ckm}
\eea
so the bulk mass parameters for the left-handed quarks are constrained by the CKM mixings, as follows,
\bea
\varepsilon^{|\nu^1_q|-|\nu^2_q|} \sim \lambda, \quad \varepsilon^{|\nu^2_q|-|\nu^3_q|} \sim \lambda^2, \quad \varepsilon^{|\nu^1_q|-|\nu^3_q|} \sim \lambda^3
\eea
where $\lambda\simeq 0.22$ is the Cabibbo angle. Therefore, from eqs.~(\ref{umass}), (\ref{dmass}) and (\ref{ckm}), the bulk mass parameters for the right-handed quarks are constrained to satisfy the following relations,
\bea
\varepsilon^{|\nu^i_u|-|\nu^j_u|} &=&(V^{ij}_{\rm CKM})^{-1} \frac{m^i_u}{m^j_u},\quad  i<j, \\
\varepsilon^{|\nu^i_d|-|\nu^j_d|} &=&  (V^{ij}_{\rm CKM})^{-1}\frac{m^i_d}{m^j_d}, \quad  i<j.
\eea
Concretely, using the ratios of quark masses \cite{pdg},
\bea
\frac{m_u}{m_c}&\sim & \lambda^{4.2} , \quad \frac{m_c}{m_t} \sim \lambda^{3.2} , \quad \frac{m_u}{m_t} \sim\lambda^{7.5}, \\
\frac{m_d}{m_s}&\sim& \lambda^2 ,\quad  \frac{m_s}{m_b} \sim \lambda^{2.5}, \quad \frac{m_d}{m_b} \sim\lambda^{4.5},
\eea
we have
\bea
\varepsilon^{|\nu^1_u|-|\nu^2_u|} &\sim&\lambda^{3.2},  \quad \varepsilon^{|\nu^2_u|-|\nu^3_u|} \sim\lambda^{1.2},  \quad \varepsilon^{|\nu^1_u|-|\nu^3_u|} \sim\lambda^{4.5}, \\ 
\varepsilon^{|\nu^1_d|-|\nu^2_d|} &\sim&\lambda,  \quad \varepsilon^{|\nu^2_d|-|\nu^3_d|} \sim\lambda^{0.5},  \quad \varepsilon^{|\nu^1_d|-|\nu^3_d|} \sim\lambda^{1.5}.
\eea
Therefore, for instance, for $\varepsilon\sim \lambda$ as in the case with a large $M_5$ from Fig.~\ref{epsilon}, we can explain the hierarchy of quark masses and mixings for the ${\cal O}(1)$ differences in the bulk mass parameters, without a fine-tuning. 
However, if $\varepsilon\ll \lambda$, which is the case with a small $M_5$, we would need to fine-tune the bulk mass parameters to get the right quark masses and mixings.

In summary we can fix nine of the total eleven parameters, $\nu^i_q, \nu^i_u, \nu^i_d$ and $y_u, y_d$, from six quark masses and the CKM mixings as above, up to the quark CP phase, so there are two free parameters unfixed, namely, $\nu^3_q, \nu^3_u$.

\subsection{Charged leptons and neutrino masses}

Regarding the flavor structure of the leptons, there are a variety of options in our model, depending on the Majorana mass terms for the right-handed neutrinos. 

When the brane-localized Majorana mass terms for the right-handed neutrinos are nonzero and larger than the Dirac neutrino masses, we can obtain the small Majorana neutrino masses by see-saw mechanism, with suppressed Dirac neutrino masses due to the localization of zero modes of bulk fermions, so the effective Majorana neutrino masses for the right-handed neutrinos, that is, $M'_R$, can be much lower than in see-saw mechanism in four dimensions. When $M'_R$ is comparable to or larger than the KK masses, which are of order $k$, the level mixings between KK modes of the bulk neutrinos on the branes would be important \cite{rsnu}. But, we ignore the level mixings in following discussion by assuming that $M'_R<k$. 

On the other hand, when the brane-localized Majorana mass terms for the right-handed neutrinos vanish by the accidental lepton symmetry, that is, $M_R=0$, we can also get the small Dirac neutrino masses and mixing angles from the localization of zero modes of bulk fermions.

Similarly to the case with quarks, assuming that the brane-localized Yukawa couplings for leptons, $y_e, y_\nu$, are flavor-diagonal, that is, $y^{ij}_e=y_e\, \delta_{ij}$, $y^{ij}_\nu=y_\nu\, \delta_{ij}$, we want to generate the realistic flavor structure for charged leptons and neutrino oscillations from the localization of the zero modes of bulk fermions. 
In the presence of a nonzero $M^{ij}_R$, if flavor-diagonal, that is, $M^{ij}_R=M_R\,\delta_{ij}$, we would get $M^{\prime ij}_R=M_R\, \eta^i_n \eta^j_n$, leading to a massless right-handed fermion, which is not relevant for see-saw mechanism.
So, in order to keep three right-handed neutrinos massive, we need to take $M^{ij}_R$ to deviate from being flavor diagonal. 

Then, after electroweak symmetry breaking, from eq.~(\ref{Lfzero}), the mass matrices for charged leptons are given by
\bea
M^{ij}_e &=& \frac{1}{\sqrt{2}}\, k y_e\,v \, \eta^i_{l} \eta^j_{e}.
\eea
Then, assuming that $\eta^1_\psi<\eta^2_\psi<\eta^3_\psi$ with $\psi=l,e$, we can diagonalize the charged mass matrices by the bi-unitary transformations \cite{hmlee,rotation}, 
\bea
V_{e_L} M_e V^\dagger_{e_R} &=& {\rm diag}(m_e,m_\mu,m_\tau)\equiv m^i_e\, \delta_{ij}
\eea
where
\bea
V^{ij}_{e_L} &\sim& {\rm min}\Big(\frac{\eta^i_l}{\eta^j_l}, \frac{\eta^j_l}{\eta^i_l}\Big)\sim  {\rm min}\Big(\varepsilon^{|\nu^i_l|-|\nu^j_l|},\varepsilon^{|\nu^j_l|-|\nu^i_l|}\Big), \\
V^{ij}_{e_R} &\sim& {\rm min}\Big(\frac{\eta^i_e}{\eta^j_e}, \frac{\eta^j_e}{\eta^i_e}\Big)\sim  {\rm min}\Big(\varepsilon^{|\nu^i_e|-|\nu^j_e|},\varepsilon^{|\nu^j_e|-|\nu^i_e|} \Big),
\eea
and $m^i_e=k y_e\, v \,\eta^i_l\eta^i_e$.
Thus, for $ky_e\sim 1$ and $\eta^3_l\eta^3_e\sim \frac{m_\tau}{m_t}\sim 10^{-2}$,  we obtain the correct tau lepton mass. 
Moreover, the mass hierarchies for charged leptons are given by
\bea
\frac{m^i_e}{m^j_e}&=& \varepsilon^{|\nu^i_l|-|\nu^j_l|}\,\times\,  \varepsilon^{|\nu^i_e|-|\nu^j_e|}, \quad i<j, \label{clmass} 
\eea
For instance, for $|\nu^i_l|=|\nu^j_l|$ for all $i,j$, using the charged lepton masses \cite{pdg}, we can determine the mass parameters for right-handed charged leptons in powers of the Cabibbo angle as
\bea
 \varepsilon^{|\nu^1_e|-|\nu^2_e|}\sim \lambda^{3.5}, \quad \varepsilon^{|\nu^2_e|-|\nu^3_e|}\sim \lambda^{6.4}, \quad \varepsilon^{|\nu^1_e|-|\nu^3_e|}\sim \lambda^{9.9}. \label{leptonmasses}
\eea
Therefore, for $\varepsilon\sim \lambda$, similarly to the case with quark masses, we can explain the hierarchy of lepton masses for the mild differences in the bulk mass parameters for leptons.

On the other hand, for $M_R\neq 0$, the Majorana masses for active neutrinos are generated by see-saw mechanism as
\bea
M_\nu = -M_D (M'_R)^{-1} M^T_D.
\eea
with $M^{ij}_D= \frac{1}{\sqrt{2}}\,k y_\nu\, v\, \eta^i_l \eta^j_n$ and $M^{\prime ij}_R=k\, M^{ij}_R\, \eta^i_n\eta^j_n$.
Then, for $M_R \eta_n\gg \eta_l y_\nu v$, the standard see-saw mechanism gives rise to small neutrino masses, $m^i_\nu\sim \frac{(\eta^i_l k y_\nu v)^2}{kM_R}$ for $\eta^i_l\ll 1$, even for $k y_\nu={\cal O}(1)$ and a relatively small $k M_R$. 

On the other hand, if $M_R=0$, the active neutrinos have only Dirac masses, which are given by
\bea
M^{ij}_\nu =M^{ij}_D = \frac{1}{\sqrt{2}}\,k y_\nu\, v\, \eta^i_l \eta^j_n.
\eea
In  this case, we can achieve small neutrino masses for $\eta^i_l \eta^i_n\ll 1$, even for $y_\nu={\cal O}(1)$. 
As $\eta^i_l $ are constrained by the masses of the charged leptons from eq.~(\ref{leptonmasses}), we can take $\eta^i_n\lesssim 6\times 10^{-9}$ from $m_\nu\lesssim 0.1\,{\rm eV}$ and $\eta^i_l\lesssim 10^{-4}$.
In this case, we would need the expansion parameter to be $\varepsilon\sim 10^{-9}$, which can be achieved being compatible with a low $M_5$ to solve the hierarchy problem as shown in Fig~\ref{epsilon}.

Therefore, after diagonalizing the neutrino mass matrix in either case by 
\bea
V_{\nu_L} M_\nu  V^T_{\nu_L} = {\rm diag}(m_1,m_2,m_3)\equiv m^i_\nu\, \delta_{ij}, \quad M_R\neq 0,
\eea
or
\bea
V_{\nu_L} M_\nu  V^\dagger_{n_R} = {\rm diag}(m_1,m_2,m_3)\equiv m^i_\nu\, \delta_{ij}, \quad M_R=0,
\eea
we can obtain both realistic masses and mixings for neutrino oscillations.
Finally, in either cases with or without $M_R$, the PMNS matrix is also obtained as
\bea
V^{ij}_{\rm PMNS}= (V_{e_L}V^\dagger_{\nu_L}  )_{ij}
\label{pmns}
\eea
so the bulk mass parameters for  leptons are constrained by the PMNS mixings.
As the mixings for the charged leptons are naturally suppressed by $\varepsilon^{|\nu^i_l|-|\nu^j_l|}$ with $i<j$ for $|\nu^i_l|>|\nu^j_l|$, the mixing angles in the PMNS matrix are determined mainly by $V^\dagger_{\nu_L}$, which depends on the neutrino mass matrix $M_\nu$.  In principle, we can obtain the realistic PMNS matrix from neutrino data \cite{NuFit,t2k} by choosing the bulk mass parameters for leptons appropriately, but we don't go to the details on the phenomenological discussion any further in this work.

In summary, for Majorana neutrinos, there are  twelve parameters in total, $\nu^i_l, \nu^i_e, \nu^i_n$ and $y_e, y_\nu, M_R$, eight of  which are fixed from three charged lepton masses, $\Delta m^2_{12}$ and $\Delta m^2_{23}$ for neutrino masses, and three neutrino mixing angles, $\theta_{12}, \theta_{23}, \theta_{13}$, up to the leptonic CP phase, so there are four free parameters unfixed, that is, $\nu^3_l, \nu^3_e, \nu^3_n$ and $M_R$.
On the other hand, for Dirac neutrinos with $M_R=0$, there are eleven parameter in total, so there are three free parameters unfixed, that is, $\nu^3_l, \nu^3_e, \nu^3_n$.

\subsection{KK gauge boson couplings}

Taking $a<0$, we can obtain the sizable gauge couplings for the zero modes of the SM gauge bosons from $g_0=g_{5D}\sqrt{k|a|/(1-e^{-2k|a| z_c})}$ in eq.~(\ref{zero-gauge}), which become $g_0\approx g_{5D}\sqrt{k|a|}$.  In this case, from eqs.~(\ref{delocal}) and (\ref{local}), for $c>0$, we get the couplings of the charged SM fermions to the KK modes of SM gauge bosons  approximately as
\bea
  g_{\psi,n}\approx \frac{ g_{5D} \pi n}{z^{3/2}_c m_{A_n}}\,\frac{4 |-3c\pm 2\nu_\psi| (3a-3c\pm 2\nu_\psi) (kz_c)^2}{9n^2\pi^2+ (3a-6c\pm 4\nu_\psi)^2 (k z_c)^2} \left\{\begin{array}{c} (-1)^{n+1} e^{k|a|z_c}, \,\, |\nu_\psi|<\frac{3}{2}c, \vspace{0.3cm}\\ 1, \quad |\nu_\psi|>\frac{3}{2}c,  \end{array} \right. \label{kkgauge-approx}
\eea
where  $3a+2(-3c\pm 2\nu_\psi)>0$ is also assumed in the latter case and $g_{5D}=g_{Y,5D}, g_{L,5D}, g_{S,5D}$ for the SM hypercharge, weak gauge bosons and gluons, respectively.
For $c<0$, we can interchange the conditions on the bulk mass parameters in eq.~(\ref{kkgauge-approx}).

As a result, for the zero modes of bulk fermions (light quarks and leptons) localized towards the brane at $z=z_c$,  have enhanced couplings to  the KK modes of SM gauge bosons as compared to those for the zero-mode gauge bosons, unless $k z_c$ is of order one or $|a|\lesssim 1/(kz_c)$. On the other hand, the zero modes of bulk fermions (top and/or bottom quarks) localized towards the brane at $z=0$ have similar couplings to the KK modes of SM gauge bosons, as compared to those for the zero-mode gauge bosons.
Therefore, the KK gauge bosons would be produced copiously from the fusion of light quark and anti-quark at the LHC, and the dijet resonance researches for the KK gauge bosons are promising. 

We remark that as noted in the previous sections, for $|a|\lesssim 1/(kz_c)$, the similar couplings for all the zero modes of bulk fermions to the KK modes of SM gauge bosons are obtained, independently of the localizations, but parametrically smaller than those for the zero-mode gauge bosons. However, from $m_{A_n}=\sqrt{a^2 k^2+ \pi^2 n^2/z^2_c}\sim \pi n/z_c$,  the masses of the KK gauge bosons become parametrically smaller than the KK masses for bulk gravitons or fermions, which are about $k$.

\subsection{KK graviton couplings}

We can parametrize the effective couplings to the SM particles in the following form \cite{GMDM},
\bea
{\cal L}_{\rm eff} &=& \frac{c_{1,n}}{\Lambda} G^{(n)\mu\nu} \Big( \frac{1}{4} \eta_{\mu\nu} B_{\lambda\rho} B^{\lambda\rho}+B_{\mu\lambda} B^\lambda\,_{\nu} \Big)+\frac{ c_{2,n}}{\Lambda} G^{\mu\nu}  \Big( \frac{1}{4} \eta_{\mu\nu} W_{\lambda\rho} W^{\lambda\rho}+W_{\mu\lambda} W^\lambda\,_{\nu}  \Big)  \nonumber \\
&&+ \frac{c_{3,n}}{\Lambda} G^{\mu\nu}  \Big( \frac{1}{4} \eta_{\mu\nu} g_{\lambda\rho} g^{\lambda\rho}+g_{\mu\lambda} g^\lambda\,_{\nu} \Big) -\frac{ic_{\psi,n}}{2\Lambda}G^{\mu\nu}\left(\bar{\psi}\gamma_{\mu}\overleftrightarrow{D}_{\nu}\psi-\eta_{\mu\nu}\bar{\psi}\gamma_{\rho}\overleftrightarrow{D}^{\rho}\psi\right) \nonumber \\
&&+\frac{c_{H,n}}{\Lambda}G^{\mu\nu}\left(2(D_{\mu}H)^{\dagger}D_{\nu}H-\eta_{\mu\nu}\left((D_{\rho}H)^{\dagger}D^{\rho}H-V(H)\right)\right).
\eea
Here, we note that the effective coupling of the KK graviton is normalized to
\bea
\Lambda=M^{3/2}_5 \sqrt{z_c}\,\frac{m_{G_1} z_c}{\pi}\simeq M_P\,e^{-kz_c}\, \frac{( kz_c)^{3/2}}{\pi}
\eea
where we used $m_{G_1}\simeq k$ for $kz_c\gg 1$ and eq.~(\ref{Pmass-CW}), and assumed $e^{k z_c}\gg1$.

Consequently, we obtain the partial decay rates of the first KK graviton $G\equiv G^{(1)}$ \cite{GMDM}, as follows,
\bea
    \Gamma_G(gg)&=& \frac{c_{gg}^2m_G^3}{10\pi \Lambda^2}, \quad\quad
       \Gamma_G(\gamma\gamma)=\frac{c_{\gamma\gamma}^2m_G^3}{80\pi \Lambda^2},  \nonumber \\
      \Gamma_G(ZZ)&=&\frac{m_G^3}{80\pi \Lambda^2}\sqrt{1-4r_Z}\bigg(c_{ZZ}^2+\frac{c_H^2}{12}+\frac{r_Z}{3}\left(3c_H^2-20c_Hc_{ZZ}-9c_{ZZ}^2\right) \nonumber \\
    &+&\frac{2r_Z^2}{3}\left(7c_H^2+10c_Hc_{ZZ}+9c_{ZZ}^2\right) \bigg),   \nonumber  \\\Gamma_G(WW)&=&  \frac{m_G^3}{40\pi \Lambda^2}\sqrt{1-4r_W}\bigg(c_{WW}^2+\frac{c_H^2}{12}+\frac{r_W}{3}\left(3c_H^2-20c_Hc_{WW}-9c_{WW}^2\right) \nonumber \\
    &+&\frac{2r_W^2}{3}\left(7c_H^2+10c_Hc_{WW}+9c_{WW}^2\right)\bigg),   \nonumber \\
    \Gamma_G(Z\gamma)&=& \frac{c_{Z\gamma}^2m_G^3}{40\pi \Lambda^2}(1-r_Z)^3\left(1+\frac{r_Z}{2}+\frac{r_Z^2}{6}\right),   \nonumber 
\eea
\bea
    \Gamma_G(\psi\bar{\psi})&=&\frac{N_c c_\psi^2 m_G^3}{160\pi \Lambda^2} (1-4r_\psi)^{3/2}(1+8r_\psi/3),  \nonumber \\
      \Gamma_G(hh)&=&\frac{c_H^2m_G^3}{960\pi \Lambda^2}(1-4r_h)^{5/2}
\eea
where $c_{\gamma\gamma}=s_{\theta}^2c_2+c_{\theta}^2c_1$, $c_{ZZ}=c_{\theta}^2c_2+s_{\theta}^2c_1$, $c_{Z\gamma}=s_{\theta}c_{\theta}(c_2-c_1)$, $c_{gg}=c_3$, $c_{WW}=2c_2$, with $c_1\equiv c_{1,0}$, $c_2\equiv c_{2,0}$,  etc, $r_i=(m_i/m_G)^2$, and $m_G=m_1$ is the lightest KK graviton mass.  

On the other hand, the decay rate of the $n$th KK graviton $G^{(n)}$  into a gluon pair becomes
\bea
    \Gamma_{G^{(n)}}(gg)=  \frac{n^2 m^2_{G_1}}{m^2_{G_n}}\, \cdot \frac{c_{gg}^2m_{G_n}^3}{10\pi \Lambda^2}=\frac{n^2 m_{G_n}}{m_{G_1}}\cdot \Gamma_{G},
\eea
etc.  The overall factor, $\frac{n^2m_{G_n}}{m_{G_1}}$, is approximated to $n^2$ for $k z_c\gg 1$,  so the partial decay widths of heavier KK gravitons are larger than the one for the first KK graviton.

For the realistic masses and mixings for quarks and leptons, heavy quarks such as top and/or bottom quarks tend to be localized towards the brane at $z=0$, whereas light quarks and leptons are localized towards the brane at $z=z_c$. Therefore, the KK gravitons can decay sizably into a pair of top or bottom quarks. On the other hand, the SM gauge bosons propagate into the bulk, so the couplings between the zero modes of transverse SM gauge bosons and the KK gravitons would have a mild suppression. But, as the Higgs field is localized on the brane $z=0$, the couplings between the longitudinal components of $W$ and $Z$ bosons and the KK gravitons are unsuppressed.

From the general results in eqs.~(\ref{graviton-bulk}) and (\ref{graviton-brane}), we can get the effective couplings of the KK gravitons to  the Higgs fields, the transverse polarizations of SM gauge bosons, and top and bottom quarks, as follows,
\bea
c_{H,n} &=&\frac{n m_{G_1}}{m_{G_n}},  \\
c_{i,n} &=&\frac{n m_{G_1}}{m_{G_n}}\, \frac{4|a|(1-a)(k z_c)^2}{|e^{2k a z_c}-1|} \cdot \frac{\Big(1-(-1)^n\, e^{(2a-1)k z_c}\Big)}{n^2\pi^2+ (1-2a)^2 (k z_c)^2},\qquad i=1,2,3,  \\
c_{q_{3L},n}
&\approx &\frac{n m_{G_1}}{m_{G_n}}\,\frac{4|-3c+ 2\nu_{q_{3L}}|(3-3c+ 2\nu_{q_{3L}}) (kz_c)^2}{9n^2\pi^2+ (3-6c+ 4\nu_{q_{3L}})^2(k z_c)^2}, \\
c_{f_{R},n} 
&\approx &\frac{n m_{G_1}}{m_{G_n}}\,\frac{4|-3c- 2\nu_{f_R}|(3-3c- 2\nu_{f_R}) (kz_c)^2}{9n^2\pi^2+ (3-6c- 4\nu_{f_R})^2(k z_c)^2},
\eea
with $f_R=t_R, b_R$.
We note that the KK graviton couplings to top and bottom quarks are comparable to those to the Higgs fields localized on the brane at $z=0$ for $|\nu_{q_{3L}}|,|\nu_{f_R}|>\frac{3}{2}c$ for $c>0$ or $|\nu_{q_{3L}}|,|\nu_{f_R}|<\frac{3}{2}|c|$ for $c<0$. 
Moreover, as we discussed in the previous subsection, the perturbativity of couplings of the KK gauge bosons would require a small $|a|\lesssim 1/(kz_c)$ for $e^{k z_c}\gg 1$. In this case, the couplings of the KK gravitons to the transverse polarizations of SM gauge bosons are parametrically smaller than those for the Higgs fields and  top and bottom quarks.
Consequently, the KK gravitons decay dominantly into the Higgs fields and  top and bottom quarks.

\section{Conclusions}

We introduced various bulk fields with general dilaton couplings in the linear dilaton background in five dimensions, and showed the bulk profile of the zero mode as well as the KK spectrum in each case. 
In particular, the localization of the zero mode of a bulk fermion depends both on  the bulk dilaton coupling and on the bulk mass parameter.
Universality of the effective coupling to massless gauge bosons determines the dilaton couplings to brane-localized matter fields while perturbativity of the effective couplings to massive gauge bosons constrains the sign and magnitude of the bulk dilaton couplings to gauge bosons. We also showed that the couplings of zero modes to the massive KK gravitons depend on the localization in the extra dimension.

Constructing the Clockwork SM in the linear dilaton background, we provided the general discussion on the effective Yukawa couplings between the zero modes of the SM fermions on the brane, and showed the necessary conditions for the bulk mass parameters for the mass hierarchy and mixing in the quark sector as well as in the lepton sector. We can introduce a sizable expansion parameter, $\varepsilon=e^{-\frac{2}{3} kz_c}$, or a small inverse proper length $L^{-1}_5$, for the realistic flavor structure in the quark sector without a fine-tuning in the bulk mass parameters, but at the expense of a large 5D Planck scale. On the other hand, we can use a smaller expansion parameter or a larger proper length of the extra dimension for realistic lepton masses, in particular, for Dirac neutrino masses, being compatible with the solution to the hierarchy problem of the Higgs mass parameter. 
We found that massive KK gauge bosons and massive KK gravitons couple more strongly to light and heavy fermions, respectively, so there is a complementarity in the resonance researches for those KK modes at the LHC.

\section*{Acknowledgments}

The work of HML is supported in part by Basic Science Research Program through the National Research Foundation of Korea (NRF) funded by the Ministry of Education, Science and Technology (NRF-2019R1A2C2003738 and NRF-2018R1A4A1025334). 
The work of YJK is supported in part by the National Research Foundation of Korea (NRF-2019-Global Ph.D. Fellowship Program).

\def\theequation{A.\arabic{equation}}

\setcounter{equation}{0}

\vskip0.8cm
\noindent
{\Large \bf Appendix A:  Matter Lagrangians with dilaton couplings}

We list the bulk and brane matter Lagrangians with or without dilaton factors in Jordan frame and show the corresponding Lagrangians in Einstein frame.

\underline{Bulk matter Lagrangian}:

For massless bulk matter fields, the corresponding kinetic terms  in Jordan frame is given by
\bea
{\cal L}_B= \sqrt{G} \,\bigg( e^{c_\chi S} G^{MN} D_M\chi D_N \chi^* + e^{c_\psi S}\, i{\bar\psi} \gamma^M D_M\psi - \frac{1}{4}  e^{a S} G^{MP} G^{NQ}F_{MN} F_{PQ} \bigg).
\eea
Under the scale transformation with $S\rightarrow S+\delta$ and $G_{MN}\rightarrow e^{-2\delta/3} G_{MN}$, bulk scalar, fermion and gauge fields transform as $\chi\rightarrow e^{-(c_\chi-1)\delta/2} \chi$, $\psi\rightarrow e^{-(c_\phi-\frac{4}{3})\delta/2} \psi$, and $A_M\rightarrow  e^{-(a-\frac{1}{3})\delta/2} A_M$.

Going to the Einstein frame with $G_{MN}=e^{-2S/3} G^E_{MN}$, the above bulk matter Lagrangian becomes
\bea
{\cal L}_B= \sqrt{G_E} \bigg(  e^{(c_\chi-1) S}\, G^{MN}_E D_M\chi D_N \chi^* +e^{c_\psi S}\, i{\bar\psi}' \gamma^M_E D_M\psi' - \frac{1}{4}\, e^{(a-\frac{1}{3})S} \, G^{MP}_E G^{NQ}_E F_{MN} F_{PQ} \bigg)
\eea
where $\gamma^M_E$ are the gamma matrices defined for the Einstein frame metric and the redefined fermion field is given by $\psi'=e^{-2S/3} \psi$.

\underline{Brane-localized matter Lagrangian}:

For massless matter fields localized on the branes, the corresponding kinetic terms in Jordan frame is given by
\bea
{\cal L}_b= \delta(z-z_i)\,\frac{\sqrt{G}}{\sqrt{-G_{55}}} \,\bigg(G^{\mu\nu} D_\mu\chi D_\nu \chi^* + i{\bar\psi} \gamma^\mu D_\mu\psi - \frac{1}{4} \, G^{\mu\rho} G^{\nu\sigma} F_{\mu\nu} F_{\rho\sigma}  \bigg).
\eea
Going to the Einstein frame with $G_{MN}=e^{-2S/3} G^E_{MN}$, the above brane matter Lagrangian becomes
\bea
{\cal L}_b= \delta(z-z_i)\,\frac{\sqrt{G_E}}{\sqrt{-G^E_{55}}} \bigg(e^{-2S/3} G^{\mu\nu}_E D_\mu\chi D_\nu \chi^* + i{\bar\psi}' \gamma^\mu_E D_\mu\psi' - \frac{1}{4} \, G^{\mu\rho}_E G^{\nu\sigma}_E F_{\mu\nu} F_{\rho\sigma} \bigg)
\eea
with the redefined fermion field being $\psi'=e^{-S/2} \psi$.

\end{document}